\documentclass{aa}  

\usepackage{graphicx}

\usepackage{txfonts}
\usepackage{amsmath}
\usepackage{amssymb}
\usepackage{placeins}
\usepackage{wrapfig}

\usepackage[colorlinks=true,linkcolor=blue,citecolor=blue,urlcolor=black]{hyperref}

\begin{document}

   \title{Red giant asteroseismic binaries in the \emph{Kepler} field}

   \subtitle{Identifying gravitationally bound systems}
   
   \author{Francisca Espinoza-Rojas\inst{1,2},
           Nathalie Themeßl\inst{2,1},
         \and
           Saskia Hekker\inst{1,2}   
          }

   \institute{Heidelberger Institut für Theoretische Studien, Schloss-Wolfsbrunnenweg 35, D-69118 Heidelberg, Germany \\
              \email{francisca.espinoza@h-its.org}, 
              \and
             Center for Astronomy (ZAH/LSW), Heidelberg University, Königstuhl 12, D-69117 Heidelberg, Germany\\
             }

   \date{Received mm dd, yyyy; accepted mm dd, yyyy}

   \abstract
   {Systems in which two oscillating stars are observed in the same light curve, so-called asteroseismic binaries (ABs), arise from either chance alignments or gravitationally bound stars. In the latter case, the detection of ABs offers a novel way to find binary systems and enables the combined use of asteroseismology and orbital dynamics to determine precise stellar parameters for both stars. Such systems provide valuable opportunities to test stellar models and calibrate asteroseismic scaling relations. While population synthesis studies predict approximately 200 ABs in the \emph{Kepler} long-cadence data, only a few have been detected to date.} 
   {Our aim is threefold. We aim to (1) expand the sample of detected ABs in \emph{Kepler} data, (2) estimate global asteroseismic parameters for both stars in each AB, and (3) assess whether these systems are gravitationally bound.}
   {We performed an asteroseismic analysis of 40 well-resolved ABs identified in the \emph{Kepler} long-cadence data. We matched these solar-like oscillators with \emph{Gaia}~DR3 sources using spectroscopic estimates of their frequency of maximum oscillation power, $\nu_{\rm max}$. To assess whether each pair is gravitationally bound, we checked their projected separation and parallax consistency, and compared the observed total orbital velocity differences derived from astrometry with theoretical predictions from Keplerian orbits.
   }   
   {Most ABs appear to be chance alignments. However, we detected two systems, KIC~6501237 and KIC~10094545, with orbital velocities, seismic masses, and evolutionary stages consistent with a wide binary configuration, i.e. they have binary probability of $\sim50$\% and $\sim25$\%, respectively. Furthermore, we found eleven ABs that are likely spatially unresolved binaries based on \emph{Gaia} multiplicity indicators.} 
   {Our findings suggest that most seismically resolved ABs in the \emph{Kepler} field are not gravitationally bound, in contrast to earlier population synthesis predictions. Remarkably, the two wide binary candidates identified here represent promising benchmarks for asteroseismic calibration. Spectroscopic follow-up is necessary to confirm their binary nature.}


   \maketitle

\section{Introduction} \label{sec:intro}

   Binary systems are essential astrophysical laboratories, offering a model-independent means of determining fundamental stellar properties, such as mass and, in the case of eclipsing systems, radius \citep[e.g.][]{Andersen1991,Torres10}. These so-called \emph{dynamical parameters} can be derived with high accuracy and precision by applying Kepler's laws, using long-term photometric time-series observations in combination with precise radial velocity measurements \citep[e.g.][]{Helminiak19}. 

   Asteroseismology --the study of the internal structure of stars through their oscillations \citep[see for example][]{Brown1994,JCD04,Aerts10}-- has proven to be a complementary method for estimating fundamental parameters with high precision. These oscillations probe internal physical properties like sound speed and density. In particular, low-mass stars exhibit solar-like oscillations, which are standing waves stochastically excited by near-surface convection and pressure gradient as the restoring force. These oscillations are primarily characterised by two global asteroseismic parameters: the frequency of maximum oscillation power, $\nu_{\rm max}$, and the large frequency separation, $\Delta\nu$. The former represents the centre of the Gaussian-shaped power excess caused by the solar-like oscillations\footnote{The frequency of maximum oscillation power $\nu_{\rm max}$  scales with the acoustic cut-off frequency \citep{Brown1991, Kjeldsen1995}, which marks the transition point where acoustic waves are trapped in a cavity or become travelling waves.}. The large frequency separation --defined as the frequency difference between pressure modes of the same degree $\ell$ and consecutive radial orders-- is proportional to the square root of the mean density \citep{Ulrich86}. When combined with spectroscopic effective temperatures, these parameters enable mass and radius estimates via scaling relations, typically achieving uncertainties below $8\%$ in mass \citep[e.g.][]{Miglio12,Yu18,APOKASC3} and a few per cent in radius \citep[e.g.][]{Huber12,White13, Huber17,Wang23}. However, it is important to note that uncertainties can be underestimated when scaling relations are calibrated with stellar models \citep[see][]{Betrisey24}.
 
   Thanks to space-based missions such as \emph{Kepler}, which continuously monitored the same field for over four years \citep{Borucki10,Howell14}, precise estimates of mass and radius are now available for around 16,000 red-giant stars \citep[e.g.][]{Pinsonneault18,Yu18,APOKASC3}. 
    
   Photometric and asteroseismic data have facilitated the detection and characterisation of eclipsing and eccentric binary systems. For instance, approximately three dozen eclipsing binaries with solar-like oscillators have been identified through their characteristic dips in brightness \citep[e.g.][]{Hekker10, Gaulme13, Gaulme14,Themessl18a, Benbakoura21}. In addition, a subset of highly eccentric, mostly non-eclipsing binaries with red-giant components has been detected via ellipsoidal modulations observed during periastron passage \citep[e.g.][]{Beck14, Beck15}. Furthermore, asteroseismic data has also been useful to characterise spectroscopic binaries. Recently, \cite{Beck22} identified 99 systems with a red-giant primary based on the ninth catalogue of spectroscopic binary orbits \citep{Pourbaix04}. Many of these systems were also observed by space-based missions like \emph{Kepler}, K2 \citep[e.g.][]{Howell14}, TESS \citep[Transiting Exoplanet Survey Satellite,][]{Ricker14}, and BRITE \citep[BRIght Target Explorer, e.g.][]{Weiss14}. In these systems, asteroseismic analysis of the high-quality photometric data enabled precise determinations of stellar mass, radius, and evolutionary stage, supporting ensemble studies of red-giant binaries and their tidal interactions. 
   
\begin{figure}
      \centering
      \includegraphics[width=0.49\textwidth]{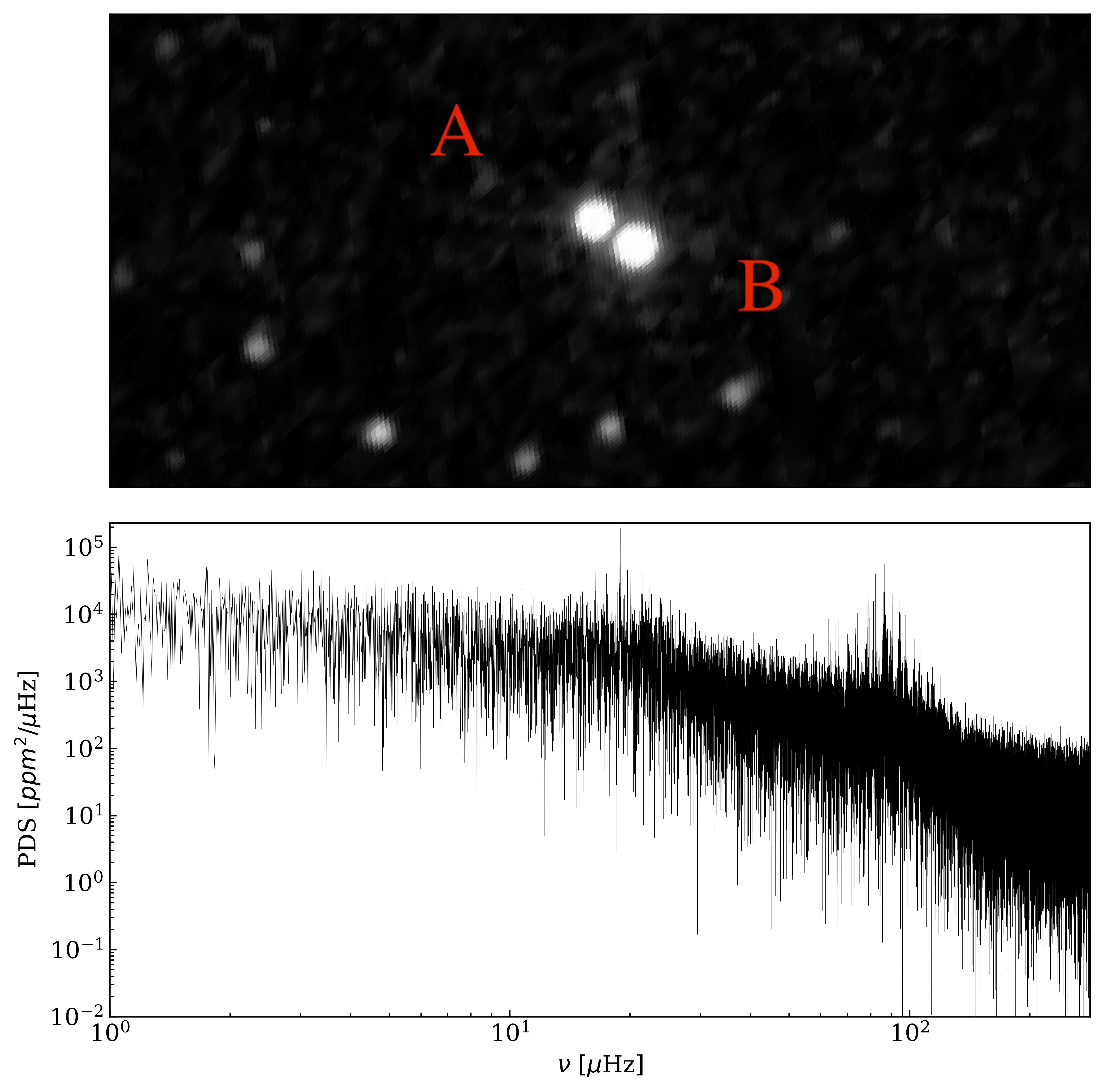}
         \caption{Image of the asteroseismic binary KIC~8004637 from the Two Micron All Sky Survey in the JHK$_{\text{s}}$ bands (top panel) and its Fourier spectrum from \emph{Kepler} data (bottom panel). Given the small spatial separation of these stars in the sky, their oscillations are detected in the same light curve.}\label{fig:AB}
   \end{figure}

   The analysis of \emph{Kepler} data has also led to the discovery of another type of binaries: the asteroseismic binaries (ABs). These systems consist of two oscillating stars located so close on the sky that they are observed within a single light curve (see Fig.~\ref{fig:AB} for an example). Hence, an asteroseismic binary exhibits two sets of stellar oscillations in the frequency spectrum, which can result from either chance alignments or gravitationally bound systems \citep[e.g.][]{Miglio14,Themessl18b}. 
   
   A theoretical binary population synthesis study by \cite{Miglio14} predicted the presence of at least 200 asteroseismic binaries in the \emph{Kepler} long-cadence dataset. The majority of these systems are expected to be composed of two core helium-burning (CHeB) red-clump stars, with orbital periods, $P_{\rm orb}$, ranging from $10$ to $10^{8}$ days.  More recently, \cite{Mazzi25} indicated that asteroseismic binaries are expected to exhibit a mass ratio close to 1 and are unlikely to interact (e.g. via mass transfer). Such interactions could hinder the detection of oscillations from both components, which favours a preserved initial semi-major axis.
   
   Given that \emph{Kepler} observed approximately 20,000 red-giants in long-cadence mode, this prediction corresponds to an occurrence rate of roughly $1$ per cent, highlighting the rarity of such systems. \cite{Miglio14} estimated a false-positive rate (i.e. spatially coincident but unassociated stars that appear within the 4-arcsecond \emph{Kepler} pixel) of less than 10\%. This implies that most of the detectable ABs should arise from genuine, gravitationally bound systems. Their detection is therefore particularly valuable, as it provides the opportunity to combine two independent techniques --asteroseismology and orbital dynamics-- to infer and cross-validate stellar properties with high precision. Direct comparisons between seismic and dynamical estimates of stellar mass and radius make these systems fundamental benchmarks for testing and calibrating asteroseismic scaling relations. Although there are numerous binary systems in the \emph{Kepler} field comprising at least one solar-like oscillator, only five confirmed gravitationally bound ABs have been reported to date \citep[e.g.][]{Appourchaux15, Rawls16, White17, Beck18, Li18}, all comprising main-sequence or sub-giant stars.

   Asteroseismic binaries comprised by two CHeB stars are likely to have similar oscillation frequencies due to their narrow range in $\nu_{\rm max}$. This similarity often results in them being seismically unresolved, complicating their detection. \cite{Choi25} demonstrated that Shannon entropy, a concept from communication theory \citep{Shannon1948}, can be used to identify these configurations. However, conducting a systematic search for these systems remains a challenge. In contrast, asteroseismic binaries comprising at least one red-giant branch (RGB) star are more likely to be seismically resolved and thus easier to detect. Although they represent only a small fraction of the predicted asteroseismic binary population, they provide a valuable initial sample for testing predictions regarding binary populations. 

   In this study, we aim to identify asteroseismic binaries formed by gravitationally bound stars. \citet{Miglio14} predicted that approximately 80 per cent of the ABs should have orbital periods longer than $10^{3}$ days. Such systems can therefore be expected to be formed by widely separated stars. 
   
   Wide binaries, also referred to as common proper motion pairs, are characterised by semi-major axes exceeding 100~AU and orbital periods ranging from $10$ to $10^{8}$~yr. Although easily resolvable on the sky, their wide separations make their detection susceptible to contamination from chance alignments, therefore precise astrometric measurements are necessary to confirm gravitational binding. For this reason, we apply a detection method based on Kepler's third law and \emph{Gaia} data (including parallaxes, proper motions, and radial velocities) to analyse a sample of 40 \emph{Kepler} targets whose Fourier spectra exhibit two distinct sets of solar-like oscillations within frequency ranges characteristic of red-giant stars. In particular, we follow the work of \cite{ElBadry21}, in which they constructed the largest catalogue of \emph{bona fide} wide binaries from \emph{Gaia}~DR3 to date based on their observed total orbital velocity differences.

   This paper is organised as follows. In Section~\ref{sec:data}, we describe the sample selection, and in Section~\ref{sec:taco} we present the asteroseismic analysis of \emph{Kepler} light curves. We provide details on the cross-matching between \emph{Gaia} DR3 and \emph{Kepler} sources in Section~\ref{sec:surveys_xmatch}, followed by the assessment of gravitational binding in Section~\ref{sec:grav_bound}. In Section~\ref{sec:results} we discuss the binary likelihood of two wide binary candidates, KIC 10094545 and KIC 6501237, along with new candidates for spatially unresolved systems. Finally, we summarise our findings in Section~\ref{sec:summary}. 
   \begin{figure*}
      \includegraphics[width=\textwidth]{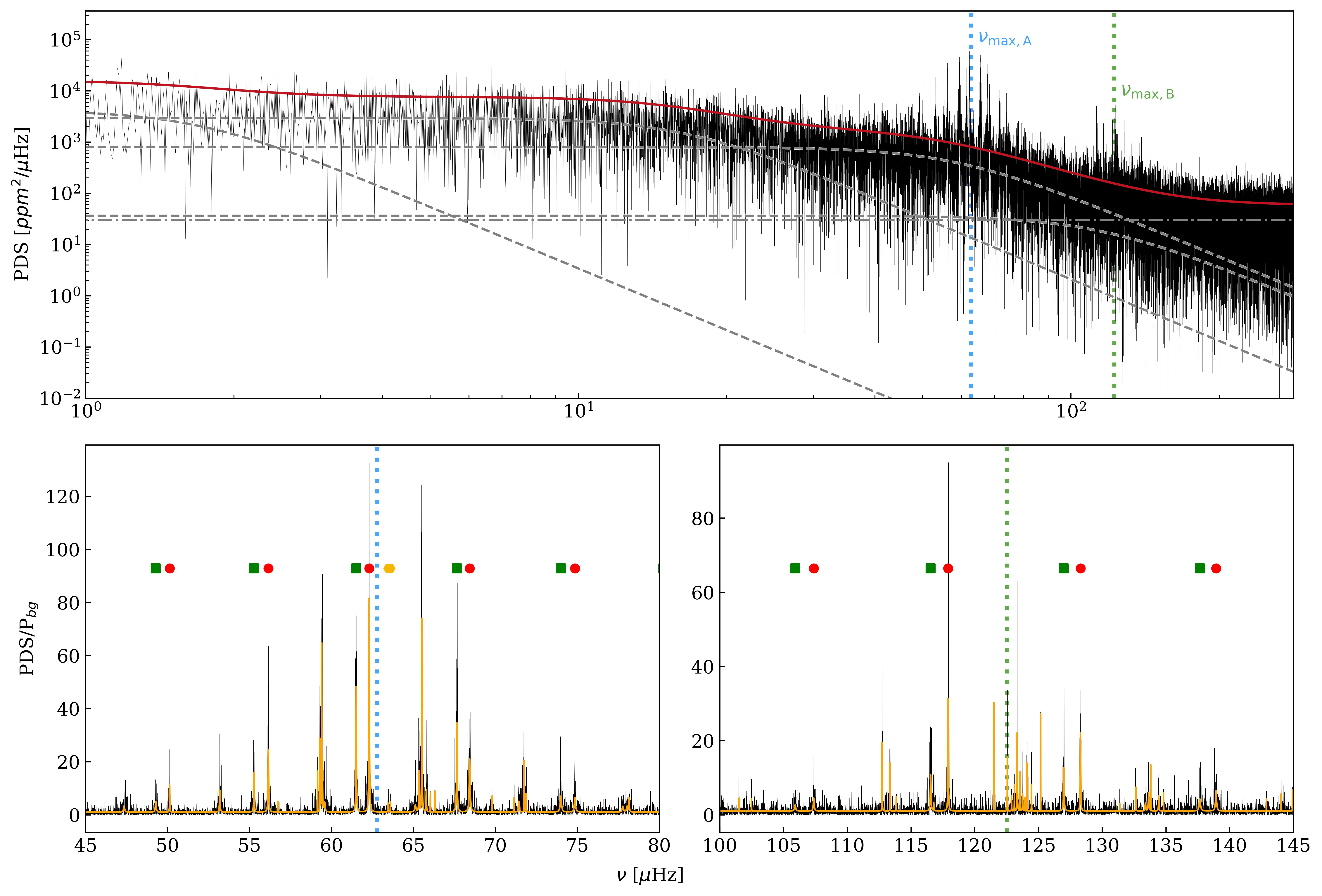}
        \caption{Power density spectrum and peak-bagging results of KIC 6501237. In the top panel, the blue and green dotted lines indicate $\nu_{\text{max, A}}$ and $\nu_{\text{max, B}}$, respectively. The four granulation components are represented by dashed grey lines, while the white noise is shown as a dot-dashed line. The total background model is depicted as a solid red line. The lower panels show the fitted oscillation modes of both solar-like oscillators in orange. Radial, quadrupole, and octupole modes are marked by red circles, green squares and yellow hexagons.}\label{fig:pds_taco}
   \end{figure*}

\section{Sample of asteroseismic binaries} \label{sec:data}

   Our sample comprises 44 asteroseismic binaries observed in \emph{Kepler} long-cadence ($\Delta t = 29.4$ min) data. Of these, 27 systems were identified by \cite{Bell2019} using pre-processed light curves from KASOC\footnote{\url{https://kasoc.phys.au.dk/}}. These systems exhibit multiple power excesses within the same power density spectrum (PDS), and were detected via the Coefficients of Variation (CV) method. This method, specifically designed for long-cadence \emph{Kepler} data, assumes that the PDS derived from the Lomb-Scargle periodogram of a Gaussian white noise time-series follows a chi-squared distribution with two degrees of freedom ($\chi^{2}_{2}$). Under this assumption, the coefficient of variation--defined as the ratio of the standard deviation to the mean of the power--should equal 1. Any signal not attributable to stochastic noise, such as solar-like oscillations, will exhibit higher CV values. 

   The remaining 17 systems were identified serendipitously during the preparation of CAPASS (Catalogue of Asteroseismic Parameters for the Analysis of Stellar Structure, Espinoza-Rojas et al. in prep), using light curves from 
   KEPSEISMIC\footnote{\url{https://archive.stsci.edu/prepds/kepseismic/index.html}}
   \citep{Garcia11}. These targets were initially flagged due to significant discrepancies between their $\nu_{\rm max}$ estimates derived in our work and those reported in the literature \citep[e.g.][]{Yu18, Pinsonneault18, Kallinger19}. Upon inspection, these discrepancies were found to result from the presence of two solar-like oscillators. Although the CV method is not applicable to KEPSEISMIC data due to altered noise statistics introduced during the data reduction process \citep[see][for further details]{Garcia11, Garcia14, Pires15}, these visually identified systems complement the KASOC sample.
   
   There are five ABs detected in CAPASS that also have KASOC light curves. However, they were not reported by \cite{Bell2019}. After revising these systems using the CV method, we confirmed the presence of two solar-like oscillators in KIC~7345204, KIC~7697607, and KIC~10094545. In the case of KIC~4663623, the star at lower frequencies is significantly affected by photometric dilution, which prevents it from meeting the detection threshold adopted by \cite{Bell2019}, despite its oscillations being visible in the power density spectrum. Furthermore, the KASOC light curve of KIC~7966761 reveals only the low-frequency oscillator, whereas the higher-frequency companion is detectable exclusively in the KEPSEISMIC light curve. Accordingly, we used KASOC data for the first four asteroseismic binaries and KEPSEISMIC for the latter. 

   Finally, we identified four duplicated asteroseismic binaries, i.e.  two \emph{Kepler} targets exhibiting the same pair of oscillations: KIC~4663623/KIC~4663627, KIC~5024297/KIC~5024312, KIC~9893437/KIC~9893440, and KIC~11090673/KIC~11090674. After removing duplicates, our final sample consists of 40 unique seismically resolved asteroseismic binaries.

\section{Asteroseismic analysis}\label{sec:taco}
   The power density spectrum of a single red-giant star consists of a granulation background and a power excess hump produced by solar-like oscillations. In contrast, our sample of asteroseismic binaries exhibits PDS with two distinct sets of oscillations, each well separated and confined to different frequency range (see top panel of Fig.~\ref{fig:pds_taco} as an example). To extract their global asteroseismic parameters, we used the data-driven code Tools for Automated Characterisation of Oscillations (TACO, Hekker et al. in prep). Given the dual nature of our targets, we modified TACO to account for the presence of two power excesses in the PDS rather than just one. 
   
   For clarity, throughout this paper we refer to the component with lower-frequency oscillations as the primary, or “star A”, and the higher-frequency component as the secondary, or “star B”. 
   
   \subsection{Background and $\nu_{\rm max}$ fit}\label{subsec:bgr_fit}

      We employed a modified version of the global fitting approach by \cite{Kallinger14} to model the PDS. The model can be described by a superposition of granulation, $P_{\rm gran}$, power excesses from stellar oscillations, $P_{\rm osc}$, of both stars, and a white noise term $W_{\rm noise}$,
      \begin{equation}\label{eq:PDS_model}
         P(\nu) = W_{\rm noise} + \eta^{2}(\nu)[P_{\rm gran}(\nu) + P_{\rm osc}],
      \end{equation}
      where $\eta = {\rm sinc}(\pi\nu/2\nu_{\rm Nyq})$ is the apodisation accounting for discrete flux sampling, and  $\nu_{\rm Nyq}$ is the Nyquist frequency.

      The granulation background is modelled using a sum of $s$ super-Lorentzian components:
      \begin{equation}
         P_{\rm gran} = \sum_{i=1}^{s} \frac{A_{i}}{1 + (\nu/b_{i})^{4}},
      \end{equation}
      where $A_{i}$ and $b_{i}$ are the characteristic amplitude and frequency, respectively. While two or three components are typically sufficient to characterise the granulation background, we used $s=4$ to account for additional granulation noise introduced by the second star. 
      
      Each power excess hump is modelled as a Gaussian envelope centred on the respective frequency of maximum oscillation power $\nu_{\rm max, i}$,
      \begin{equation}\label{eq:Posc}
         P_{\rm osc}(\nu) = \sum_{i=A,B} P_{\rm g, i} \exp \bigg(-\frac{(\nu - \nu_{\rm max, i})}{2 \sigma_{\rm env, i}^{2}} \bigg),
      \end{equation}
      where $P_{\rm g, i}$ and $\sigma_{\rm env, i}$ are the height and width of the Gaussian envelope of star $i$. 
      
      The fitting of the PDS model given by Eq.~\ref{eq:PDS_model} is performed using an affine-invariant Markov Chain Monte Carlo (MCMC) algorithm \citep{MCMC10} implemented via the {\tt emcee} Python package \citep{emcee}. This implies that the variables of the Gaussian envelope (i.e. $\nu_{\rm max}, P_{g}$ and $\sigma_{{\rm env}}$), granulation components (i.e. $A_{i}$, $b_{i}$), and W$_{\rm noise}$ are free parameters. We took the median values of the posterior distributions as our best-fit parameters, with uncertainties defined by their 16th and 84th percentiles. An example fit for KIC~6501237 is shown in Fig.~\ref{fig:pds_taco}, where the vertical dotted blue and green lines indicate the fitted $\nu_{\rm max}$ values of the two components, located at around 62.7~$\mu$Hz and 122.5~$\mu$Hz.

      To normalise the PDS, we divided the power by the global model fit excluding the two Gaussian components $P_{\rm osc}$ in Eq.~\ref{eq:PDS_model} (see solid red line in Fig.~\ref{fig:pds_taco}). Subsequently, we extracted individual normalised PDS within the frequency range $\nu_{\rm max} \pm 3\sigma_{\text{env}}$ of each star. These background-normalised spectra were used for the remainder of the analysis.

   \subsection{Estimation of the large frequency separation $\Delta\nu$}\label{subsec:dnu_estimate}
      The identification of the spherical degree $\ell$ of each oscillation mode is essential for determining various asteroseismic parameters. In particular, radial modes ($\ell=0$) are used to estimate the large frequency separation $\Delta\nu$, which is the spacing between consecutive overtones of the same degree. To estimate $\Delta\nu$, we performed a linear regression of the identified $\ell=0$ modes as a function of radial order to compute a global value of $\Delta\nu$, and assigned the standard error as uncertainty. Further details regarding this process and the mode identification of $\ell=0, 2$, and 3 are provided in Appendix~\ref{subsec:appendix_modeID}. The lower panels of Fig.~\ref{fig:pds_taco} illustrate the peak fitting and mode identification results for both stars in KIC~6501237.

   \subsection{Fundamental stellar properties from scaling relations}\label{subsec:fundamental_params}    
      Asteroseismic scaling relations were originally developed to predict the frequencies of the solar-like oscillators from spectroscopic atmospheric parameters \citep[see][for a review]{Hekker20a}.
      \cite{Brown1991} empirically showed that $\nu_{\rm max}$ scales with the acoustic cut-off frequency $\nu_{\text{ac}}$, which represents a typical atmospheric dynamical timescale and is therefore proportional to the stellar surface gravity $g$. This relation is commonly expressed as \citep{Kjeldsen1995}:    
      \begin{align}\label{eq:spec_numax}
         \frac{\nu_{\rm max}}{\nu_{\rm max, ref}} &\simeq \bigg(\frac{g}{g_{\odot}} \bigg) \bigg(\frac{T_{\rm eff}}{T_{\rm eff, ref}} \bigg)^{-1/2} \\
         &\simeq \bigg(\frac{M}{M_{\odot}}\bigg) \bigg(\frac{R}{R_{\odot}}\bigg)^{-2} \bigg(\frac{T_{\rm eff}}{T_{\rm eff, ref}}\bigg)^{-1/2}  .
      \end{align}
     
      In addition, \cite{Kjeldsen1995} showed that $\Delta\nu$ is proportional to the square root of the mean stellar density $\bar{\rho}$, based on the asymptotic approximation. This relation is also supported by detailed stellar models \citep{Ulrich86}, and can be expressed as: 
      \begin{equation}
         \frac{\Delta\nu}{\Delta\nu_{\rm ref}} \simeq \bigg(\frac{M}{M_{\odot}}\bigg)^{1/2} \bigg(\frac{R}{R_{\odot}}\bigg)^{-3/2}.
      \end{equation}
      Here “ref” denotes reference values. Following \citet{Themessl18a}, we used $\nu_{\rm max, ref}= 3166\pm 6$ $\mu$Hz, which is derived from various solar estimates, and $\Delta\nu_{\rm ref} = 130.8\pm 0.9$ $\mu$Hz to account for structural differences between the Sun and red-giant stars. We also adopted $T_{\rm eff, \odot} = 5771.8\pm0.7$~K \citep{Mamajek15,Prsa16}. By rearranging the above expressions, stellar mass and radius can be estimated with the following scaling relations:
      \begin{equation}
            \frac{M}{M_{\odot}} \simeq \bigg(\frac{\nu_{\rm max}}{\nu_{\rm max, ref}} \bigg)^{3} \bigg( \frac{\Delta\nu}{\Delta\nu_{\rm ref}} \bigg)^{-4} \bigg(\frac{T_{\rm eff}}{T_{\rm eff,ref}}\bigg)^{3/2},
      \end{equation}\label{eq:seismic_mass_SR}
         \begin{equation}
            \frac{R}{R_{\odot}} \simeq \bigg(\frac{\nu_{\rm max}}{\nu_{\rm max, ref}} \bigg) \bigg( \frac{\Delta\nu}{\Delta\nu_{\rm ref}} \bigg)^{-2} \bigg(\frac{T_{\rm eff}}{T_{\rm eff,ref}}\bigg)^{1/2}.
      \end{equation}

      Several studies have proposed refinements to these scaling relations to improve the accuracy of the stellar parameter determinations, accounting for deviations due to structural differences between red giants and the Sun, surface effects, and dependencies on metallicity or stellar mass \citep[e.g.][]{Sharma16, Miglio16, Pinsonneault18,Themessl18a, Hekker20a}. In this work, we apply corrections only through the use of the $\Delta\nu$ reference value calibrated by \cite{Themessl18a}, which incorporates mass, temperature, metallicity dependence as well as surface effects. This calibration was shown to improve the agreement between asteroseismic and dynamical estimates of stellar parameters. Since our goal is to obtain only approximate estimates of stellar mass and radius, we consider this calibration sufficient and do not apply further adjustments to the scaling relations.
      
      In addition to estimating fundamental parameters, we also use the scaling relations for their original purpose: predicting the oscillation frequencies of solar-like stars. This approach is especially useful given the relatively large pixel size of \emph{Kepler}, which complicates the source matching of our sample of red giants with \emph{Gaia}~DR3. Asteroseismic parameters, particularly $\nu_{\rm max}$, provide an additional constraint in this process. We defined $\nu_{\rm max}^{\text{spec}}$ as the frequency of maximum oscillation power predicted from surface gravity and effective temperature (Eq.~\ref{eq:spec_numax}), and used it in our source-matching procedure (see Sect.~\ref{subsec:source_identification}).

\section{Identification of asteroseismic binary members} \label{sec:surveys_xmatch}
   Identifying \emph{Gaia} DR3 counterparts to individual stars in asteroseismic binaries is challenging due to the limited spatial resolution of \emph{Kepler}. Within a single Target Pixel File (TPF), multiple nearby stars can fall within the photometric aperture and potentially contribute their light to the structure which is visible in the PDS. We refer to these possible contributors as AB-member candidates, and we aim to identify which of them are the observed solar-like oscillators.
    
   We used the {\tt interactive\_sky} function provided by the {\tt lightkurve} package\footnote{\url{https://lightkurve.github.io/lightkurve/index.html}} \citep{lightkurve} to extract the {\tt source\_id} of each candidate. This tool displays \emph{Gaia} sources within the field of view of each \emph{Kepler} TPF. Given the similarity between the \emph{Gaia} $G$-band and the \emph{Kepler} $K_{\rm p}$-band, we limited our selection to sources with $7 < \texttt{phot\_g\_mean\_mag} < 17$, ensuring sensitivity to the most likely contributors to the observed signal. This process yielded 101 AB-member candidates across our sample, with an average of $2.5$ AB-member candidates per TPF.

   \subsection{Astrometric and spectroscopic parameters from surveys}\label{subsec:survey_data}
      We extracted astrometric and photometric parameters from \emph{Gaia}~DR3 for all AB-member candidates, including positions ($\alpha, \delta$), proper motions ($\mu$), parallaxes ($\varpi$), $G$-band photometry, and, when available, radial velocities ($v_{r}$). To assess potential unresolved multiple systems, we also retrieved their Renormalised Unit Weight Error (RUWE) values and {\tt non\_single\_star} flags. A RUWE value around 1.0 typically indicates that the astrometric solution is consistent with a single-star model, while values exceeding 1.4 \citep[or 1.2 in less stringent cases,][]{Castro-Ginard24} suggest possible binarity \citep[e.g.][]{Lindgren18, Belokurov20, Penoyre22, Castro-Ginard24}. The {\tt non\_single\_star} flag further classifies stars as either (1) astrometric, (2) spectroscopic or (3) eclipsing binaries \citep{Halbwachs23,Gosset25}. Among the 101 AB-member candidates, 11 exhibit high RUWE values, and two\footnote{Gaia DR3 2052444444184805376 and Gaia DR3 2075400563350022784} are classified as spectroscopic binaries based on their {\tt non\_single\_star} flag. Furthermore, we inspected the {\tt phot\_variable\_class} in \emph{Gaia}~DR3 to exclude classical pulsators (e.g. RR-Lyrae, $\delta$ Scuti, Cepheids) since their oscillations do not correspond to the signals observed in the PDS of any of the asteroseismic binaries in our sample.
 
      To correct for known biases in \emph{Gaia} parallaxes, we applied the zero-point corrections described by \citet{Lindegren21a}, using the publicly available {\tt gaiadr3\_zeropoint}
      package\footnote{\url{https://gitlab.com/icc-ub/public/gaiadr3_zeropoint}}.
      In addition, we followed the approach of \citet{ElBadry21} to adjust the reported parallax uncertainties, accounting for underestimation in bright stars ($11 < G < 13$), close pairs, and sources with high {\tt RUWE} values or poor \emph{Gaia} image parameter diagnostics (e.g. {\tt ipd\_gof\_harmonic\_amplitude}~$>0.1$, {\tt ipd\_frac\_multi\_peak}~$>10$). The applied corrections have average values of $\sim 0.025$~mas in parallax and $\sim 0.041$~mas in parallax uncertainty. Furthermore, we retained only stars with good-quality astrometric data, that is with parallax uncertainties $\sigma_{\varpi} < 2$~mas and parallax fractional uncertainties $\varpi/\sigma_{\varpi} > 5$ \citep[e.g.]{ElBadry18,ElBadry21}
   
      In addition to the astrometry, we retrieved stellar atmospheric parameters derived from low-resolution BP/RP spectra from \emph{Gaia}~DR3 GSP-Phot \citep[GSP-Phot;][]{Andrae23}, and from combined radial velocity spectra from GSP-Spec \citep{Recio-Blanco23}. We also incorporated spectroscopic and photometric information from SDSS-IV/APOGEE~DR16 \citep{APOGEE2}.

   \subsection{Removing main-sequence and white-dwarf stars}\label{subsec:CMD}

      To restrict our sample of AB-member candidates to red-giant stars, we used the \emph{Gaia} colour–magnitude diagram (CMD) to exclude sources located in the main-sequence and white dwarf regions. For this purpose, we adopted the \emph{Gaia} CMD boundaries defined by \cite{GodoyRivera25}. Briefly, they used PARSEC \citep[PAdova and tRiestet Stellar Evolutionary Code;][]{Bressan12,Chen14,Nguyen22} evolutionary tracks with solar metallicity to characterise different stellar populations, including the main sequence, subgiant and red giant branch, and white dwarfs. In addition, \citet{GodoyRivera25} provides $G$-band absolute magnitudes ($M_G$) and $BP-RP$ colours corrected for extinction ($A_G$) and reddening ($E(BP-RP)$), respectively, for all the stars in the \emph{Kepler-Gaia}~DR3 cross-match\footnote{\url{https://gaia-kepler.fun}}. 
      
\begin{figure}
         \includegraphics[width=0.49\textwidth]{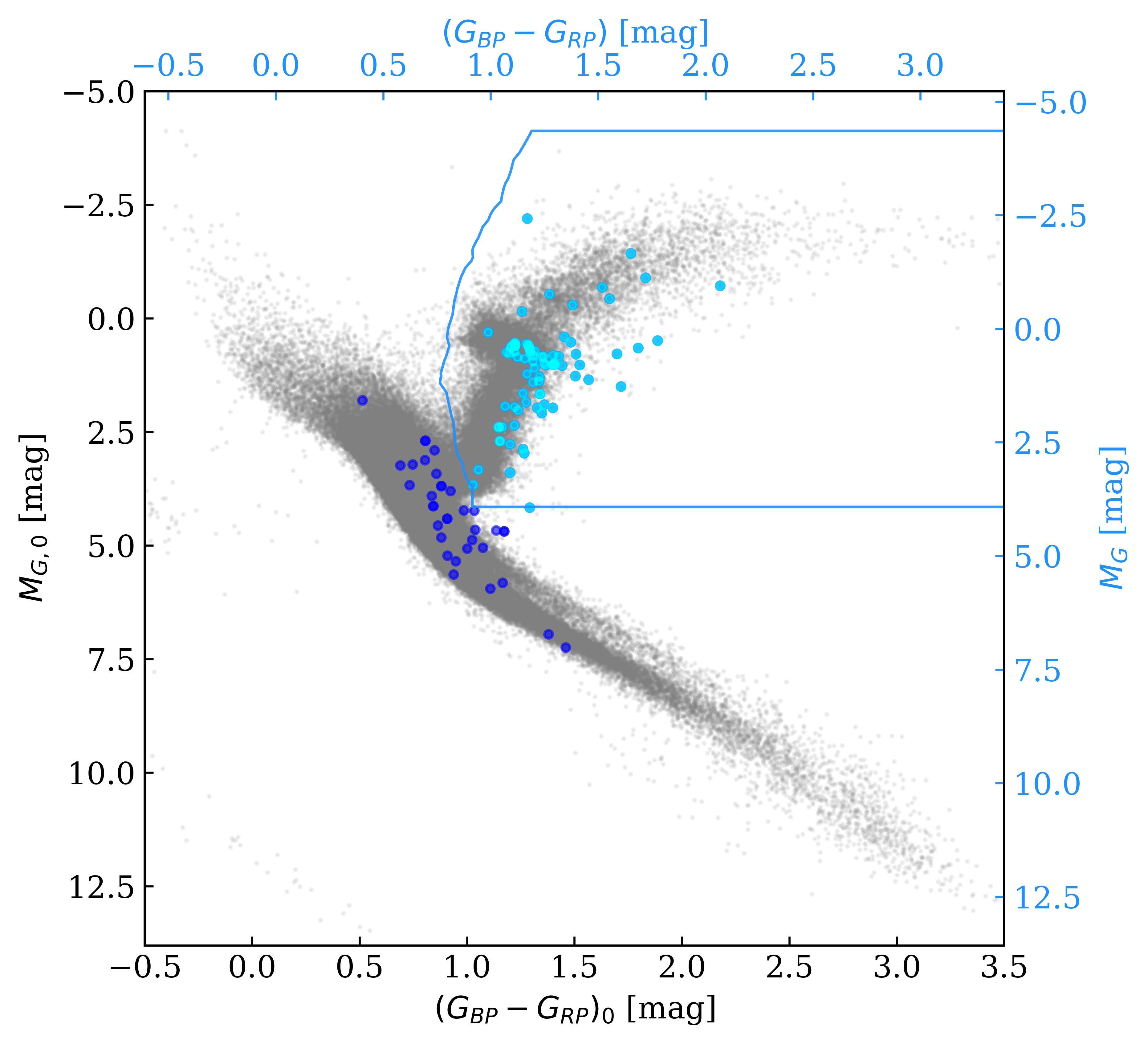}
         \caption{\emph{Gaia} colour-magnitude diagram of the AB-member candidates. Grey points represent \emph{Kepler} stars from \citet{GodoyRivera25}, with extinction- and reddening-corrected photometry, plotted using the black axes. The light blue axes are used to plot the \emph{Gaia}~DR3 photometry of our sample that lacks $E(BP-RP)$ and $A_{G}$ corrections. The light blue line depicts the borders of the giant-stars region as defined by \citet{GodoyRivera25}, adjusted to account for the missing photometric corrections. AB-member candidates classified as giants according to this criterion are shown in cyan, while those excluded by the criteria are shown as blue circles.} \label{fig:CMD}
      \end{figure}

      Since many of our AB-member candidates are not part of the \emph{Kepler}-\emph{Gaia} catalogue, extinction and reddening estimates were unavailable for a large fraction of them. While GSP-Phot provides such values, its strong distance prior introduces systematic overestimates in $T_{\rm eff}$ and $\log g$ \citep[see][]{Andrae23}. As an alternative, we adopted the mean extinction and reddening values of red giants in APOKASC~3 \citep{APOKASC3} that are part of the sample of \citet{GodoyRivera25}. We found average values of $E(BP-RP) \sim 0.11$ and $A_G \sim 0.23$. Rather than correcting each star, we applied a global shift to the red-giant region boundaries in the CMD. This adjusted region is shown as the light blue contour in Fig.~\ref{fig:CMD}. Since this filtering process requires an absolute magnitude estimate, AB-member candidates lacking \emph{Gaia} photometry or parallax estimates were also excluded.
      As discussed by \cite{GodoyRivera25}, restricting the composition of the models used to define the CMD region to solar metallicity can affect the final classification. They found that, at lower metallicities, the regions become redder and less luminous. However, they demonstrated that their classification method is robust against changes in metallicity of around $\pm$0.2~dex. We consider their approach sufficiently accurate for the purposes of our study.
       
      This filtering yielded a final sample of 71 red-giant AB-member candidates, shown as cyan circles in Fig.~\ref{fig:CMD}, which are used for the rest of the analysis. Given the small magnitude of the extinction and reddening corrections, we do not expect this approximation to significantly affect the results of our selection.

   \subsection{Matching asteroseismic and spectroscopic $\nu_{\rm max}$ values} \label{subsec:source_identification}
   For each red-giant AB-member candidate, we estimated the expected frequency of maximum oscillation power, denoted as \(\nu_{\rm max}^{\rm spec}\), using the relation \(\nu_{\rm max} \propto g / \sqrt{T_{\rm eff}}\), as described in Sect.~\ref{subsec:fundamental_params}. We prioritised spectroscopic atmospheric parameters from available surveys in the following order of spectral resolution and precision: APOGEE, GSP-Spec, and GSP-Phot.

   Atmospheric parameters from GSP-Phot, particularly effective temperature and surface gravity, are known to suffer from systematic biases. \citet{Andrae23} report typical overestimations of approximately $\delta T_{\rm eff} = 400$~K and $\delta \log g = 0.4$~dex relative to APOGEE~DR16, primarily due to strong distance priors based on $d \sim 1/\varpi$. These biases are especially pronounced for stars with low-quality parallaxes ($\varpi/\sigma_{\varpi} < 20$), which is common among distant red giants where the parallax–distance relation breaks down \citep[see][]{BJ21}. 

   Among our red giant AB-member candidates, 20 stars have atmospheric parameters from GSP-Phot only. To mitigate the aforementioned biases, we applied empirical corrections following the values reported in \citet{Andrae23}. For stars with $\varpi/\sigma_{\varpi} \leq 20$, we subtracted the median absolute deviation (MAD) from the reported $T_{\rm eff}$; for $\log g$, the MAD correction was applied only when $\varpi/\sigma_{\varpi} \leq 10$. For AB-member candidates with higher-quality parallaxes, we used the smaller Median Absolute Deviation from the Median (MedAD) values, which are considered more appropriate in the low-bias regime. The correction values were taken from Table~1 and Table~2 of \citet{Andrae23}. Applying these corrections reduces $\nu_{\rm max}^{\rm spec}$ by $10^{-\delta \log g}(1-\delta T_{\rm eff}/T_{\rm eff})^{-1/2}$ and reduces the seismic mass by $(1-\delta T_{\rm eff}/T_{\rm eff})^{3/2}$. Additionally, we inflated the reported uncertainties by factors of 2.0 for $T_{\rm eff}$ and 2.5 for $\log g$ to account for their underestimation, as recommended by \citet{Andrae23}. No corrections were applied to APOGEE (42 stars) or GSP-Spec (9 stars) parameters, as their calibrations are considered reliable within the context of this analysis.

   To identify the most likely \emph{Gaia} source for each red giant, we searched for candidates whose $\nu_{\rm max}^{\rm spec}$ fell within the range $\nu_{\rm max} \pm 2\Delta\nu$, also accounting for uncertainties in both parameters. Using this criterion, we successfully matched both solar-like oscillators to unique \emph{Gaia} sources in 15 AB systems (see Table~\ref{table:Full_identification_summary}). In 19 systems, we could confidently identify only one component (see Table~\ref{table:single_identification_summary}). Ten of these 19 cases either lacked complete spectroscopic data for all AB-member candidates or did not have $\nu_{\rm max}^{\rm spec}$ consistent with the asteroseismic estimates, i.e. $\nu_{\rm max}^{\rm spec}$ is outside the search range (see central panel of Fig.~\ref{fig:matching_results}). Additionally, we found five of these 19 systems to be associated with a single AB-member candidate that shows signs of being a non-single source (e.g. high {\tt RUWE}), suggesting possible unresolved multiplicity. These systems are discussed in Sect.~\ref{subsec:gaia_multiplicity}. In contrast, four other systems also have a single AB-member candidate and show no indication or multiplicity.

\begin{figure}
      \centering
      \includegraphics[width=0.45\textwidth]{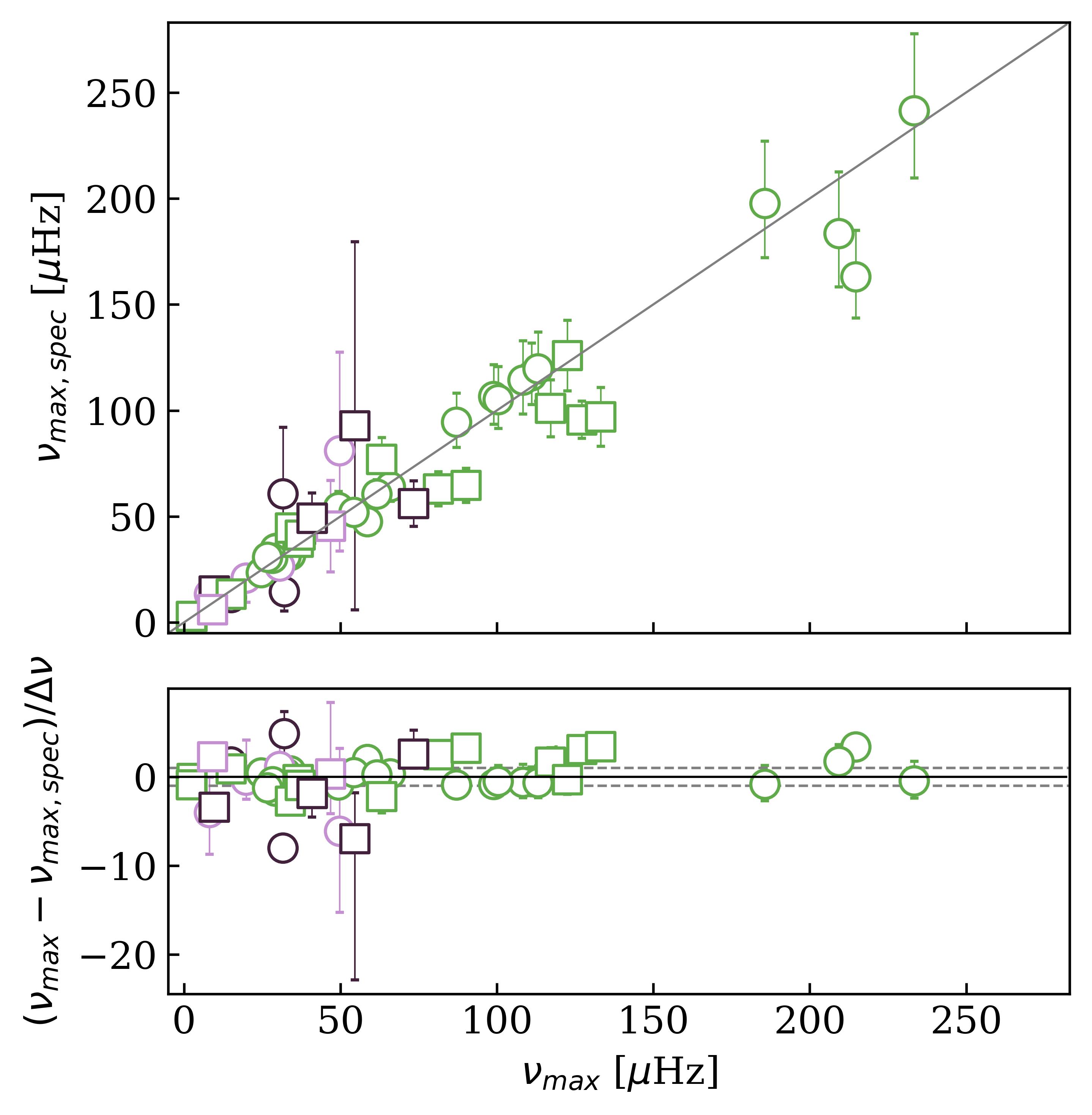}
         \caption {Comparison of asteroseismic and spectroscopic $\nu_{\rm max}$ estimates of the \emph{Kepler-Gaia} matched sources described in Sect.~\ref{subsec:source_identification}. Stars with atmospheric parameters from APOGEE, GSP-Spec, and GSP-Phot are shown in green, purple, and pink, respectively. Circles correspond to ABs where both solar-like oscillators of an AB were matched to a \emph{Gaia} source. Squares depict ABs where only one of the two oscillators was matched. The bottom panel displays the difference in units of radial order, i.e. normalised by $\Delta\nu$. Grey dotted lines represent differences of $\pm1\Delta\nu$.}\label{fig:numax_comparison}
   \end{figure}

   For the remaining six ABs, no match could be obtained for either stars, since the estimated $\nu_{\rm max}^{\rm spec}$ did not fall within the search range, or there was only one AB-member candidate for the system, which in turn had a $\nu_{\rm max}^{\rm spec}$ value between the two asteroseismic $\nu_{\rm max}$ estimates (bottom panel Fig.~\ref{fig:matching_results}). These cases are of particular interest and are further discussed in Sect~\ref{subsec:single_non_match}. 
   Fig.~\ref{fig:numax_comparison} illustrates the differences in $\nu_{\rm max}$ and $\nu_{\rm max}^{\rm spec}$ with the colours representing the source of the atmospheric parameters. As expected, spectroscopic estimates based on APOGEE data show better agreement with asteroseismic values than those from GSP-Phot or GSP-Spec.    

\section{Finding gravitationally bound stars}\label{sec:grav_bound}
   Our sample of red-giant AB-member candidates includes, on average, two \emph{Gaia} sources per \emph{Kepler} target. We computed the angular separation, $\theta$, using their \emph{Gaia} positions:
   \begin{equation}
      \theta \simeq \sqrt{(\alpha_{1} - \alpha_{2})^{2}\cos\delta_{1}\cos{\delta_{2}} + (\delta_{1} - \delta_{2})^{2}},
     \end{equation}
   where $\alpha_{i}$ and $\delta_{i}$ are the right ascension and declination, respectively. The subindex $i=1,2$ corresponds to the primary and secondary component where, by definition, the primary is the brightest star of the pair. This convention differs from the asteroseismic designations “star A” and “star B” based on oscillation frequencies. 
   
   The angular separation can be converted into a projected physical separation, $s$, using a distance proxy from their trigonometric parallax, $\varpi$:
   \begin{equation}
      s = 1000\text{AU} \times \bigg(\frac{\theta}{\text{arcsec}} \bigg) \bigg(\frac{\varpi}{\text{mas}} \bigg)^{-1}.
   \end{equation}

   We used the parallax of the brightest star in the pair as representative of the system. The smallest projected separations in our sample are larger than 1000~AU, suggesting that if any of these pairs are gravitationally bound, they are wide binaries. Accordingly, to assess the gravitational binding of our candidate pairs, we followed the approach implemented by \cite{ElBadry21}.
   
   We required that the parallaxes of the two stars in each pair be consistent within uncertainty:
   \begin{equation}\label{eq:parallax_diff}
      |\varpi_{1} - \varpi_{2}| < b\sqrt{\sigma_{\varpi,1}^{2} + \sigma_{\varpi,2}^{2}} \hspace{1cm}
      b = 
      \begin{cases}
      3 & \text{if } \theta \geq 4 \\
      6 & \text{if } \theta < 4 
      \end{cases}
   \end{equation}   
   Here, $\varpi_{1}$ and $\varpi_{2}$ are the parallaxes of the primary and secondary star, respectively, with $\sigma_{\varpi_{1}}$ and $\sigma_{\varpi_{2}}$ their corresponding uncertainties. As discussed in \cite{ElBadry21}, the constant $b$ is assigned a larger value at small separations to account for two reasons: (1) underestimated parallax uncertainties at close angular separations and (2) lower contamination rate from chance alignments in that regime.

   The wider the binary system, the less likely it is to remain bound over long timescales, as its gravitational attraction weakens with increasing separation. This makes the widest binaries most susceptible to disruption by external perturbations, like Galactic tides and dynamical encounters in dense environments. Wide binaries are rarely found beyond projected separations of $\sim 1$~pc \citep[with a few exceptions, e.g.][]{Chaname04,SO11}, corresponding to orbital periods of $P_{\text{orb}} \sim 10^{8}$ yr. As discussed by \cite{Andrews17}, this limit reflects the scale at which the binding energy of the system becomes comparable to the Galactic tidal field given by the Milky Way Jacobi radius \citep{BT08}. To avoid contamination by potentially unbound pairs, we applied the following constraint:
   \begin{equation}\label{eq:max_s}
        \frac{\theta}{\text{arcsec}} \leq 206.265 \times \frac{\varpi}{\text{mas}},
   \end{equation}
   which sets a maximum separation corresponding to $s < 1$~pc. 
    
   To test whether a given pair of stars is gravitationally bound, we required that their relative motions are consistent with Keplerian orbits. Under the assumption of circular orbits the maximum total velocity difference expected for a binary according to Kepler's third law is given by:
   \begin{equation}\label{eq:total_mass}
      \Delta V_{\text{tot, max}} = \sqrt{\frac{GM_{\text{tot}}}{a}},
    \end{equation}
   where $a$ is the semi-major axis, and satisfies $a \geq s$. 

   The total velocity difference of a binary is defined as the norm of the velocity difference vector:
   \begin{equation}
      \Delta V_{\text{tot}} = \sqrt{ \Delta V_{\perp}^{2} + \Delta V_{r}^{2} },
   \end{equation}
   where $\Delta V_{\perp}$ is the tangential velocity difference and $\Delta V_{r} =  |V_{r,1} - V_{r,2}|$, i.e. the radial velocity difference. Then, any pair of stars that meets the condition $\Delta V_{\rm tot} < \Delta V_{\rm tot, max}$ is classified as a binary system. However, radial velocities are not available for all the stars in our sample. Hence, if any of the components of a pair of stars lacks radial velocity measurements, we implement a different binary condition based exclusively on their tangential velocity difference \citep[e.g.][]{Andrews17,ElBadry18,ElBadry21,Barrientos21}:
   \begin{equation}\label{eq:dvt_limit}
      \Delta V_{\perp} < \Delta V_{\text{tot,max}} + 3\sigma_{\Delta V_{\perp}}.
   \end{equation}
   Here, $\sigma_{\Delta V_{\perp}}$ accounts for uncertainties in proper motion and parallax. Inflating the threshold by $3\sigma$ makes the criteria less conservative, but it can instead favour completeness by avoiding rejection of genuine binaries due to parallax and or proper motion noise, especially for distant stars. The tangential velocity difference is computed as:
   \begin{align}
      \Delta V_{\perp} &= 4.74 \times \bigg( \frac{\Delta\mu}{\text{mas yr}} \bigg)\times \bigg( \frac{\varpi}{\text{mas}} \bigg)^{-1} \text{km s}^{-1}, \\
      \sigma_{\Delta  V_{\perp}} &= 4.74 \text{ km s}^{-1} \sqrt{\dfrac{\Delta\mu^{2}}{\varpi^{4}}\sigma^{2}_{\varpi} + \dfrac{\sigma^{2}_{\Delta\mu}}{\varpi^{2}}},
   \end{align}
   where $\Delta\mu$ and $\sigma_{\Delta\mu}$ are the total proper motion difference and its uncertainty which are defined as:
    \begin{align}
      \Delta\mu &= \sqrt{ \Delta\mu_{\alpha}^{2} + \Delta\mu_{\delta}^{2} }, \\
      \sigma_{\Delta\mu} &= \frac{1}{\Delta\mu} \sqrt{ (\sigma_{\mu_{\alpha, 1}^{*}}^{2} + \sigma_{\mu_{\alpha, 2}^{*}}^{2}) \Delta\mu_{\alpha}^{2} + (\sigma_{\mu_{\delta, 1}}^{2} + \sigma_{\mu_{\delta, 2}}^{2}) \Delta\mu_{\delta}^{2} }.
   \end{align}
   Here, $\Delta\mu_{\alpha}^{2} = (\mu_{\alpha, 1}^{*} - \mu_{\alpha, 2}^{*})^{2}$, and $\Delta\mu_{\delta}^{2} = (\mu_{\delta, 1} - \mu_{\delta, 2})^{2}$, where $\mu_{\alpha}^{*} \equiv \mu_{\alpha} \cos\delta$ accounts for projection effects in right ascension. 

\begin{figure*}
      \centering
      \begin{minipage}{\textwidth}
         \centering
         \includegraphics[width=\textwidth]{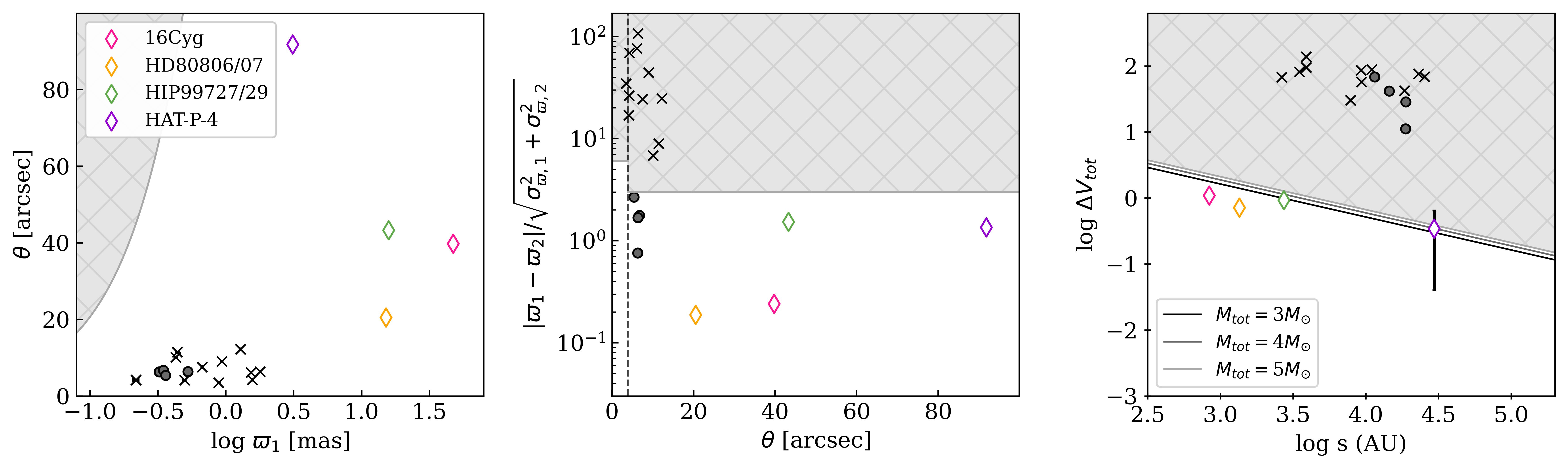}      
      \end{minipage}
      \vspace{1ex}
      \begin{minipage}{\textwidth}
         \centering
         \includegraphics[width=\linewidth]{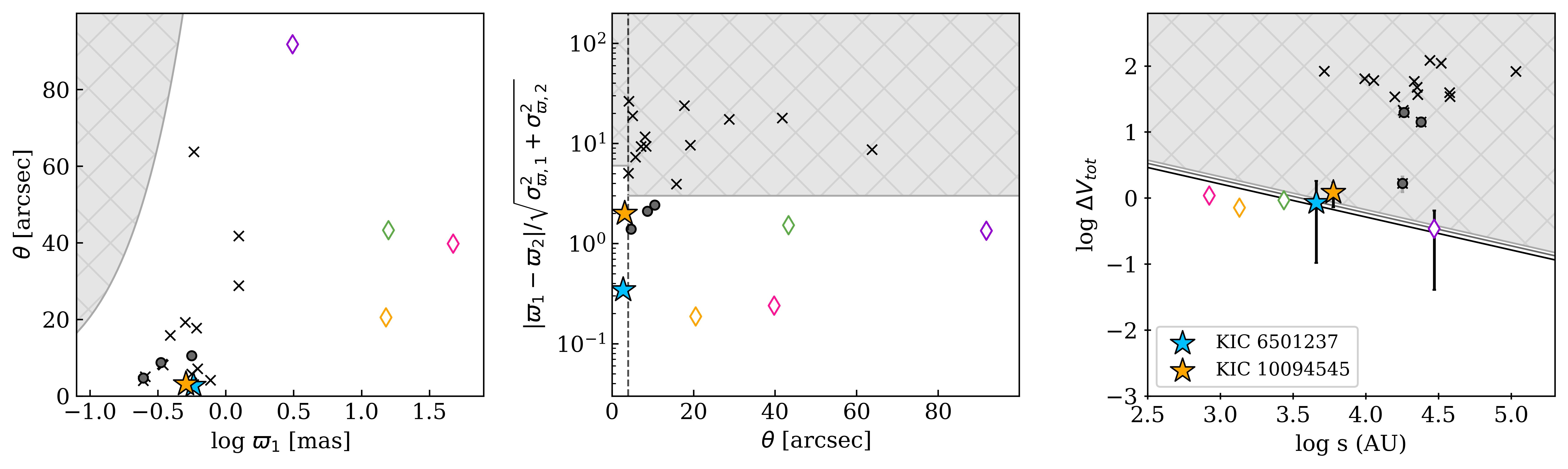}
      \end{minipage}   
      \caption{Summary of binarity criteria. Top row: AB systems with both components successfully matched to \emph{Gaia} sources (Table~\ref{table:Full_identification_summary}). Bottom row: all possible pairs from AB-member candidate sets, excluding those in the top row. Left panels illustrate angular separation as a function of parallax, with limits from Eq.~\ref{eq:max_s} ($s < 1$~pc) as grey hatched region. Centre panels show parallax consistency; black dashed lines mark the criteria  from Eq.~\ref{eq:parallax_diff}. Right panels show the total orbital velocity difference versus projected separation (log-log scale), with theoretical $\Delta V_{\text{tot}}$ values from Eq.~\ref{eq:total_mass} for M$_{\rm tot} = 3$, 4, and 5~M$_\odot$. Crosses, circles and stars represent pairs that only satisfy one, two, or three binary conditions, respectively. Well-known \emph{bona fide} wide binaries 16 Cygni, HD 80806/07, HIP 99727/29, and HAT-P-4, are shown as pink, orange, green, and purple diamonds, respectively. Genuine wide binary candidates are expected to lie within the white area of each plot. KIC~6501237 and KIC~10094545, highlighted as blue and orange stars, respectively, are identified as likely genuine wide binaries.}\label{fig:grav_bound}
   \end{figure*}

   In general, this method to detect wide binaries assumes small angular separations ($\theta < 1^{\circ}$), where curvature effects in the sky-projected coordinates can be neglected. We find that all candidate pairs in our sample have separations of $\theta < 60$~arcsec, confirming the validity of this approximation (see Fig.~\ref{fig:grav_bound}).
   
   Finally, we implemented a Monte Carlo approach to provide a probabilistic assessment of binarity accounting for uncertainties. For each pair, we generated 1000 realisations of the total velocity difference by sampling from a Gaussian distribution centred on $\Delta V_{\rm tot}$ and its corresponding uncertainty as the standard deviation. We then compared each realisation of $\Delta V_{\rm tot}$ with the theoretical velocity thresholds (see Eq.~\ref{eq:total_mass}) computed for three different total masses: 3, 4, and 5~M$_\odot$. This way, we estimated the probability of the observed motion being consistent with a bound system for each assumed total mass, yielding a set of three binary probability estimates per system. To account for biases produced by the sampling distribution selection, we repeat the same procedure using a uniform distribution centred in  $\Delta V_{\rm tot}$ and with a width given by $\pm 3\sigma_{\Delta V_{\rm tot}}$.

\section{Results}\label{sec:results}   

   \subsection{Wide binary candidates}\label{subsec:binary_candidates}

      The method described in Sect.~\ref{sec:grav_bound} was applied to two different groups: (1) asteroseismic binaries in which both oscillating components were successfully matched to unique \emph{Gaia} sources (see Table~\ref{table:Full_identification_summary}), and (2) all other possible pairs of stars within each AB-member candidate list, excluding those in group~1. For the latter, we considered every unique pair combination, ensuring that no star was paired with itself and that no duplicated combinations were included. The results for group 1 and 2 are presented on the top and bottom row of Fig.~\ref{fig:grav_bound}, respectively. 
      
      For reference, Fig.~\ref{fig:grav_bound} also includes four well-studied \emph{bona fide} wide binaries: 16~Cygni \citep{TucciMaia14}, HD~80806/07 \citep{Dommanget02,Mack16}, HIP~99727/29 \citep{vandenBergh1958,Ramirez14a}, and HAT-P-4 \citep{Mugrauer14,Saffe17}. Each of these systems is formed by solar type or solar twins stars, with estimated total mass around 2~$M_{\odot}$ \citep[see][]{Naef01,Laughlin09,Ramirez14a, Saffe17,Bellinger17}
      Using \emph{Gaia}~DR3 data, we recomputed their total velocity differences and confirmed their consistency with values previously reported in the literature \citep[e.g.][]{Ramirez19}. We also computed theoretical $\Delta V_{\text{tot}}$ values assuming total system masses of 3, 4, and 5~M$_{\odot}$, which are representative of red-giant binaries \citep[see][]{APOKASC3}. We find that this variation of $\pm 1$~M$_{\odot}$ in the total mass has minimal impact in our binary classification.

      As illustrated in Fig.~\ref{fig:grav_bound}, none of the asteroseismic binaries in group~1 appear to be gravitationally bound, as their total velocity differences significantly exceed the theoretical thresholds, even when accounting for measurement uncertainties. These discrepancies are driven by large differences in parallax, proper motion, and/or radial velocity between the components. In contrast, we identified two systems from group~2, KIC~6501237 and KIC~10094545, with kinematics consistent with a wide binary scenario. For both systems, $\Delta V_{\text{tot}}$ reaches values below the theoretical limits within $3\sigma_{\Delta V_{\text{tot}}}$ according to the total mass of each system (2.74~$M_{\odot}$ and 4.26~$M_{\odot}$, respectively. See Table~\ref{table:wb_candidates}).
       
      It is worth noting that \cite{ElBadry21} ensured the use of moderately precise astrometry by restricting their sample to stars with parallaxes greater than 1~mas, corresponding to distances of $d\lesssim1$~kpc. However, our analysis necessarily includes stars at larger distances, owing to the intrinsic brightness of red giants. As discussed in Appendix~\ref{sec:appendix_wb}, this broader distance range pushes the method to regimes where the inverse parallax could no longer be a reliable proxy for distances. To evaluate the impact of this limitation, we recalculated the total orbital velocity differences using geometric distances provided by \cite{BJ21} instead of $1/\varpi$. We find that this substitution introduces negligible differences in the estimates of $\Delta V_{\rm orb}$ and does not change the conclusions of our analysis. For a more detailed discussion, we refer the reader to Appendix~\ref{sec:appendix_wb}.
       
      \begin{table*}
         \caption{Spectroscopic, photometric and asteroseismic parameters of the wide binary candidates in our sample.} \label{table:wb_candidates}   
         \centering                     
      \begin{tabular}{cccccccccc}
         \hline\hline  
            KIC &  {\tt source\_id} & Source & $T_{\rm eff}$ & $\log g$ & Mass & Evol. stage  & $G_{\text{mag}}$ & $P_{\rm norm}$ & $P_{\rm uni}$ \\
                  &             &        &   [K]       & [dex]     &[M$_{\odot}$]  &       & [mag]  & \\
         \hline \
         6501237A  &2104510923954927104 & GSP-Phot & $5267.0^{184.6}_{94.6}$ & $3.40^{2.74}_{2.73}$ & 1.45$\pm$0.04 & RGB & 14.4  & 0.46 & 0.48\\ 
         6501237B  &2104510919655573120 &  APOGEE  & $4763.1^{84.1}_{84.1}$ & $3.00^{2.18}_{2.12}$ & 1.27$\pm$0.05 & RGB   & 12.1 &  &  \\ 
         \hline 
         10094545A &2085575547024946048 &  APOGEE  & $5027.5^{89.6}_{89.6}$  & $2.79^{1.86}_{1.81}$ & 2.52$\pm$0.15 & 2CL   & 12.1 & 0.18 & 0.35 \\ 10094545B &2085575547020615424 & GSP-Phot & $4677.4^{183.2}_{28.5}$ & $2.64^{2.16}_{1.93}$ & 1.74$\pm$0.08 & RGB   & 13.3 &  &  \\ 
         
         \hline
      \end{tabular}
      \tablefoot{ Wide binary probabilities for a total mass of 3~$\text{M}_{\odot}$ (for KIC 6501237) and 4~$\text{M}_{\odot}$ (for KIC 10094545) were computed with a normal ($P_{\rm norm}$) and uniform ($P_{\rm uni}$) distribution.}
      \end{table*}

      \subsubsection{KIC~10094545}\label{subsubsec:KIC10094545}
         
         The asteroseismic binary KIC 10094545 was identified by \cite{Bell2019} using the CV method. Its Target Pixel File contains only two AB-member candidates: Gaia~DR3~2085575547024946048 (candidate~A) and Gaia~DR3~2085575547020615424 (candidate~B). The stellar atmospheric parameters for candidate~A were taken from APOGEE, while those for candidate~B were sourced from GSP-Phot. As discussed in Sect.~\ref{subsec:source_identification}, we applied corrections to the GSP-Phot values to address known systematic errors in distant stars.
         
         Matching candidates A and B to the observed oscillations was challenging, as both candidates have spectroscopic $\nu_{\rm max}$ values within the uncertainties that agree with the value of $73\pm1\mu$Hz estimated for the primary star KIC~10094545A, i.e. $\nu_{\rm max, A}^{\rm spec}= 76_{9}^{18}\mu$Hz and $\nu_{\rm max, B}^{\rm spec}= 56_{11}^{18}\mu$Hz. We solved this by selecting the candidate with the spectroscopic estimate closest to the asteroseismic estimate, i.e. candidate~A was matched to star A. This choice is reliable given the precision of the stellar parameters from APOGEE of candidate A.
         
         However, the value of $\nu_{\rm max,B}^{\rm spec}$ remained significantly offset from the asteroseismic $\nu_{\rm max}$ of star B, which is 122~$\mu$Hz, even after applying corrections. These corrections resulted in an approximated decrease of 34~$\mu$Hz in the estimated $\nu_{\rm max, B}^{\rm spec}$\footnote{The uncorrected value was $\nu_{\rm max, B}^{\rm spec} \sim$90~$\mu$Hz}. Since we prioritised APOGEE parameters due to their more reliable calibration, this suggests that the GSP-Phot values for candidate~B can be affected by additional, uncorrected biases. Notably, the applied corrections shifted $\nu_{\rm max, B}^{\rm spec}$ in the opposite direction relative to $\nu_{\rm max,B}$, further supporting this interpretation. Consequently, we cannot confidently match the secondary star to any \emph{Gaia} source in this system. Nonetheless, given the absence of other AB-member candidates in the TPF and the astrometric consistency of candidates~A and~B, we assume from this point forward that they correspond to the primary and secondary stars observed in KIC~10094545, respectively. 
 
         KIC~10094545 consists of a secondary clump (2CL) star and a red-giant branch (RGB) star (see Appendix~\ref{sec:appendix_ES}), with inferred seismic masses of M$_{\text{A}} = 2.52 \pm 0.15$ M$_{\odot}$ and M$_{\text{B}} = 1.74 \pm 0.08$ M$_{\odot}$, respectively (see Eq.~\ref{eq:seismic_mass_SR}). Proposed wide binary formation scenarios \citep[e.g.][]{Kouwenhoven10, Tokovinin17} suggest that stars in such configurations are coeval and have similar chemical compositions \citep[e.g][]{TucciMaia14,Ramirez19,Hawkins20,ER21}. This implies that binary stars with a mass ratio different from one could be found at different evolutionary stages once they have left the main sequence, as is the case of this asteroseismic binary. While these parameters (seismic mass estimates and evolutionary stages) alone are not enough to probe the wide binary nature of the system, they provide additional evidence supporting this scenario. However, the probability that this system is gravitationally bound, as defined in Sect.~\ref{sec:grav_bound}, is around 20\% when compared to the theoretical orbital velocity difference of a 4~M$_{\odot}$ system (as shown by the orange line in Fig.~\ref{fig:grav_bound}). This probability increases to $\sim 25$\% for a total mass of 5~M$_{\odot}$.

         In summary, KIC~10094545 appears to be a promising wide binary candidate, based on the consistent asteroseismic masses and evolutionary stages of its components. However,  our astrometric analysis indicates a probability of binarity of $\sim20-25$\%, which prevents us from drawing a definitive conclusion. We encourage spectroscopic follow-up observations to better identify the secondary component and confirm binarity, for instance through radial velocity monitoring or elemental abundance analysis. To our knowledge, this is the first observational evidence supporting the binary nature of this system.
   
      \subsubsection{KIC~6501237}\label{subsubsec:KIC6501237}
      
         We successfully matched the secondary component, KIC~6501237B, to Gaia~DR3~2104510919655573120 (hereafter candidate B) 
         based on APOGEE atmospheric parameters. According to their relative astrometry, candidate~B is likely part of a wide binary with Gaia~DR3~2104510923954927104 (candidate~A). However, the atmospheric parameters for candidate~A --available only from GSP-Phot and after applying the corrections mentioned in Sect.~\ref{subsec:source_identification}-- yield $\nu_{\rm max}^{\rm spec} = 301^{+66}_{-65}~\mu$Hz, which is significantly higher than its asteroseismic value of $\nu_{\rm max,A} = 63~\mu$Hz. Similar to the case of KIC~10094545, this asteroseismic binary does not have other AB-member candidate and candidate~B does not present any \emph{Gaia} multiplicity indicators (i.e. high {\tt RUWE}, {\tt non\_single\_star} flag), therefore we assume that candidate~A corresponds to star~A, and interpret the discrepancy between $\nu_{\rm max,A}^{\rm spec}$ and $\nu_{\rm max, A}$ as large systematic biases in GSP-Phot parameters.

         KIC~6501237 was first classified as a possible close to equal mass binary by \cite{Tayar15} in the context of rapid rotation of low-mass red giants.
         Assuming candidates~A and~B form a binary system with a mass ratio close to 1, we expect them to be in similar evolutionary phases. We find that both stars in this system are on the RGB (see Table~\ref{table:wb_candidates}). Furthermore, we estimated a seismic mass of $1.27\pm0.05~M_\odot$ for candidate~B using its atmospheric parameters from APOGEE. To estimate the mass of candidate~A, we explored two approaches. First, we used its atmospheric parameters from GSP-Phot (M$_{A, \rm GSP-Phot}$). Second, we used the atmospheric parameters of candidate~B as a proxy (M$_{A,\rm APOGEE}$). These approaches yielded mass estimates of M$_{A, \rm GSP-Phot} = 1.77\pm0.04~M_\odot$ and M$_{A,\rm APOGEE} = 1.45\pm0.04~M_\odot$. On the one hand, GSP-Phot parameters lead to a relatively large mass difference, mainly due to systematically overestimated effective temperatures and surface gravities. On the other hand, the APOGEE-based result suggests a smaller discrepancy, consistent with expectations for a binary system.

         Our mass estimates are slightly higher than those reported by \cite{Tayar15}, M$_{A} = 1.31$ M$_{\odot}$ and M$_{B} = 1.39$ M$_{\odot}$, likely due to differences in $\nu_{\rm max}$, $\Delta\nu$, and reference values. While \cite{Tayar15} proposed coevality based on mass similarity, they rejected binarity due to photometric magnitude differences derived from UKIRT images. Although we were unable to confidently match the KIC~6501237A to a \emph{Gaia} source, the evidence suggests that both candidates are part of a wide binary system. In particular, the asteroseismic masses are in good agreement, supporting a coeval origin for the two stars, and their relative astrometric motions of candidates~A and B yield a binary probability of approximately 50\%. As in the case of KIC~10094545, additional high-resolution spectroscopic observations are required to confirm the \emph{Gaia–Kepler} cross-match and further establish the binary nature of the system.
   
   \subsection{KIC~2568888}\label{subsec:KIC2568888}
      We revisited KIC~2568888, an asteroseismic binary previously analysed in the study by \citet{Themessl18b}, which reported a low binary probability of only $\sim 0.1\%$ based on a combined seismic and photometric analysis.

      We identified four AB-member candidates, of which only one passed the CMD-filtering criteria described in Sect.~\ref{subsec:CMD}. This candidate\footnote{Gaia~DR3~2051291674955780992} exhibits a {\tt RUWE} value of 1.5 and has a spectroscopic $\nu_{\rm max}^{\rm spec} \sim 13 \pm 1$~$\mu$Hz, derived from APOGEE~DR17 stellar parameters, since no observations were available in APOGEE~DR16, GSP-Phot, or GSP-Spec. While this candidate was not matched to either of the oscillating components, its $\nu_{\rm max}^{\rm spec}$ lies between the two asteroseismic $\nu_{\rm max}$ values of the system (see Table~\ref{table:null_identification_summary}).
      
      We note that this APOGEE target is flagged with {\tt STARFLAG}=``PERSIST\_HIG'',
       indicating that more than 20\% of its pixels fall within the super-persistence regions of the APOGEE detectors. As discussed in \cite{Holtzman18}, this effect is wavelength dependent and can artificially enhance flux levels by over 10\%, potentially distorting spectral features. We therefore suspect that the lack of a reliable match can be due to systematics introduced by super-persistence, which can further bias the derived atmospheric parameters.

      KIC~2568888 remains an intriguing target for further study, particularly given its high {\tt RUWE} value, which suggests that it is an unresolved binary system \citep[e.g.][]{Castro-Ginard24}. However, based on the current data, we are unable to confirm or discard its binary nature.
      
   \subsection{Gaia's multiplicity indicators}\label{subsec:gaia_multiplicity}
      Radial velocity variability can indicate the presence of a close companion or, in some cases, intrinsic stellar variability. In Table~\ref{table:high_ruwe} we present a subset of seven stars that exhibit high ${\tt RUWE}$ values, implying potential multiplicity as discussed in Sect.~\ref{subsec:survey_data}. To further investigate the likelihood of these stars forming a gravitationally bound system, we examined their radial velocity variation using \emph{Gaia}~DR3 indicators. 
      
      \begin{table*}
         \caption{Red-giants in ABs with high ${\tt RUWE}$ values ($\gtrsim 1.4$).} \label{table:high_ruwe}   
         \centering                     
      \begin{tabular}{cccccc}
         \hline\hline  
         KIC  & {\tt source\_id} & ${\tt RUWE}$ & {\tt rv\_chisq\_pvalue} & {\tt rv\_renormalised\_gof} & {\tt rv\_nb\_transits} \\
                 &       &         &   &  &   \\
         \hline \
         2568888   & 2051291674955780992 & 1.50  &  -   &  -   & 10 \\ 
         
         4260884B  & 2052991687430043648 & 13.04 &  -   &  -   & 0  \\

         5443536A  & 2101197305151000832 & 1.36 &  0.0   &  55.61   & 17  \\

         6206407A  & 2053632432134403200 & 34.04 &  -   &  -   & 0  \\

         6888756A  & 2078451055276043008 & 1.81  & 0.52 & 0.19 & 13 \\
         
         7729396A  & 2116924749531060608 & 3.72  & 1e-4 & 2.73 & 12 \\
         
         10841730  & 2119742797832530176 & 47.53 & $5.2\times10^{-9}$ & 5.22 & 7 \\
         
         \hline
      \end{tabular}
      \end{table*}

      \cite{Katz23} provide two key variability indicators: {\tt rv\_chisq\_pvalue} and {\tt rv\_renormalised\_gof}. 
      These are derived from the properties of the \emph{Gaia} time series, and quantify the constancy and noise level of radial velocity measurements. Briefly, {\tt rv\_chisq\_pvalue} is expected to be small for stars exhibiting significant radial velocity scatter, as expected in binaries, while {\tt rv\_renormalised\_gof} compares the observed scatter in radial velocity to the typical epoch uncertainty. The reliability of these indices improves for stars with at least ten transits (${\tt rv\_nb\_transits} \geq 10$) and effective temperatures in the range $3900~K < {\tt rv\_template\_teff} < 8000$~K. Following the classification criteria of \cite{Katz23}, stars with ${\tt rv\_chisq\_pvalue}\leq0.01$ and ${\tt rv\_renormalised\_gof}>4$ are flagged as radial velocity variables. All stars in our sample have RV effective temperatures within the applicable range for this method. 

      As shown in Table~\ref{table:high_ruwe}, \emph{Gaia}~DR3 provides radial velocity variability indicators for only four of the seven stars in our sample. KIC~10841730 was studied in detail by \cite{Schimak23}, and was confirmed as a binary through long-term ($>4$ years) radial velocity monitoring with the HERMES spectrograph \citep{Raskin11}. The TPF of this system contains only one AB-member candidate, with a $\nu_{\rm max}^{\rm spec}$ of $\sim51\mu$Hz--located between the two asteroseismic estimates for star A and B (see Table~\ref{table:null_identification_summary}). As a result, the candidate was not matched to either component. This system thus serves as an illustrative example of an asteroseismic binary formed by a spatially unresolved binary system.
      
      We find that KIC~5443536 has only one AB-member candidate, which exhibits the strongest evidence of radial velocity variability according to the criteria of \cite{Katz23}, closely resembling the case of KIC~10841730. The combination of these variability indicators together with its high {\tt RUWE} value supports the interpretation of this system as a strong candidate of a spatially unresolved binary. Similarly, KIC~7729396A exhibits a {\tt rv\_chisq\_pvalue} of 0.0001 and a {\tt rv\_renormalised\_gof} of 2.73, indicative of significant radial velocity scatter, although these values do not fully meet the thresholds established by \citet{Katz23}. Taken together, these findings suggest that both systems are likely genuine binaries whose spatial components cannot be resolved in the current \emph{Gaia} data, and are therefore compelling targets for future spectroscopic follow-up.

      Furthermore, we identified five systems that, despite having a {\tt RUWE} and other indicators consistent with a single-star solution, show highly significant radial velocity variability (see Table~\ref{table:rv_var}). The case of KIC~8479383 stands out among them, as we could not match any of the oscillations to its single AB-member candidate; similar to the cases mentioned above, KIC 10841730 and KIC 5443536.
      
      \begin{table*}
         \caption{Red-giants in ABs with \emph{Gaia}~DR3 radial velocity variability indicators.} \label{table:rv_var}   
         \centering                     
      \begin{tabular}{cccccc}
         \hline\hline  
         KIC  & {\tt source\_id} & ${\tt RUWE}$ & {\tt rv\_chisq\_pvalue} & {\tt rv\_renormalised\_gof} & {\tt rv\_nb\_transits} \\
                 &       &         &   &  &   \\
         \hline \
         4173334B   & 2076190665528735616 & 1.0 &  2e-3  &  8.76   & 10 \\ 
         
         4663623B  & 2052444444184805376 & 1.06 &  0.0   &  24.97   & 18  \\

         6689517B  & 2101925422364578176 & 1.19 &  0.0   &  24.20   & 22  \\

         8479383  & 2106722076198209408 & 1.17  & 0.0 & 25.61 & 16 \\
         
         10592924B  & 2130946172784444544 & 1.11  & 1.2e-6 & 9.15 & 15 \\
         
         \hline
      \end{tabular}
      \end{table*}

      \begin{figure}
      \centering
      \includegraphics[width=0.45\textwidth]{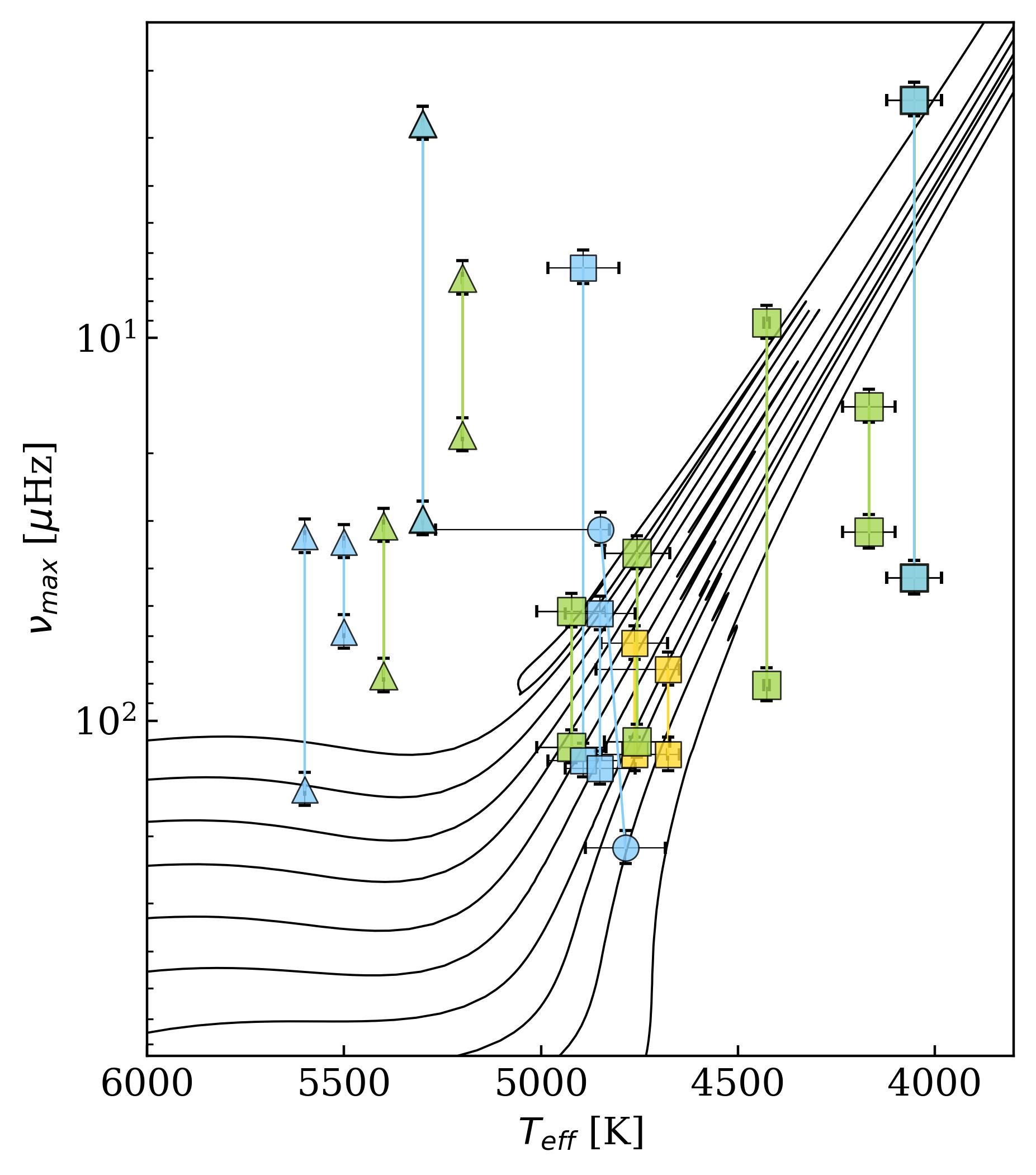}
         \caption {Asteroseismic binaries likely to be gravitationally bound systems. Colours represent the method used to detect the binarity, where yellow, green and light blue are wide binaries (Table~\ref{table:wb_candidates}), stars with high {\tt RUWE} (Table~\ref{table:high_ruwe}), and stars with RV variability indicators (Table~\ref{table:rv_var}), respectively. Circles represent ABs where both solar-like oscillators were matched to a \emph{Gaia} source (Table~\ref{table:Full_identification_summary}). Squares depict ABs where only one of the two oscillators was matched (Table~\ref{table:single_identification_summary}), and triangles correspond to ABs where none of the oscillators was matched to a \emph{Gaia} source (Table~\ref{table:null_identification_summary}). Black lines show Dartmouth evolutionary tracks \citep{Dotter08} for 0.8–2.6,M$_{\odot}$ at solar metallicity, with mass decreasing from left to right.}\label{fig:binary_candidates}

   \end{figure}
   
   \subsection{Asteroseismic binaries with no \emph{Gaia}~DR3 match}\label{subsec:single_non_match}
      As mentioned in Sect.~\ref{subsec:source_identification}, there is a subsample of six systems for which neither one of the components were matched to a \emph{Gaia} source (see Table~\ref{table:null_identification_summary}). Three of these systems, KIC~2568888, KIC~8479383, and KIC~10841739, have already been discussed (Sect.~\ref{subsec:KIC2568888} and \ref{subsec:gaia_multiplicity}). Here, we examine the remaining three cases and discuss their likelihood of being spatially unresolved binaries. 
      
      For KIC~7966761, which has only one AB-member candidate (candidate A), the absence of a match is attributed to the very low $\nu_{\rm max}$ of star~A ($2.76\pm0.04$~$\mu$Hz) combined with the relatively small uncertainty in its spectroscopic estimate  ($\nu_{\rm max, A}^{\rm spec} = 1.06^{+0.18}_{-0.14}$~$\mu$Hz) from GSP-Spec. Together, these factors result in a search range that is too narrow to allow any match. Upon re-examination of the TPF, we identified an additional \emph{Gaia} source\footnote{GaiaDR3 2078200023020874240}, candidate~A$^{*}$, previously excluded from the AB-member candidates due to missing BP and RP magnitudes (see Sect.~\ref{subsec:CMD}).

      Candidate~A$^{*}$ has an APOGEE-based spectroscopic estimate, $\nu_{\rm max, A^{}}^{\rm spec} = 2.84^{+0.35}_{-0.31}$$\mu$Hz, that closely matches the asteroseismic value of star~A, making it the most plausible counterpart. Moreover, it exhibits significant multiplicity indicators: a {\tt RUWE} of 2.08, {\tt rv\_chisq\_pvalue} of $1.3 \times 10^{-15}$, and {\tt rv\_renormalised\_gof} of 35.9 (see Sect.~\ref{subsec:gaia_multiplicity}). However, large discrepancies in parallax, radial velocity, and proper motion between candidates A and A$^{*}$ rule out any gravitational interaction between both stars. We suggest that the proximity of candidate~A$^{*}$ may have biased the GSP-Spec parameter estimates for candidate~A, thereby affecting $\nu_{\rm max, A}^{\rm spec}$ and raising the possibility that it can correspond to star~B instead. Given the strong multiplicity indicators for candidate~A$^{*}$ and its lack of association with candidate~A, its variability can arise from a companion that is likely not a red giant, as no third star is observed on the PDS of KIC~7966761. Explaining the origin of this asteroseismic binary remains a challenge and requires spectroscopic observations.

      The only AB-member candidate of KIC~6441499\footnote{\emph{Gaia}~DR3 2101724344873978752}, with a $\nu_{\rm max}$ value consistent with the red-giant branch, was excluded during the CMD-based filtering (Sect.~\ref{subsec:CMD}) due to the absence of a parallax measurement. This \emph{Gaia} source has a two-parameter solution ({\tt astrometric\_params\_solved=3}), indicating that only positions ($\alpha, \delta$) are available. Consequently, \emph{Gaia} does not provide a {\tt RUWE} value or any radial velocity variability indicators. As outlined in Sect.~4.4 of \citet{Lindegren21b}, such cases typically result from one or more astrometric parameters (or combinations thereof) being insufficiently constrained. Therefore, we are unable to draw any further conclusions regarding the nature of this asteroseismic binary.

      KIC~9412408 resembles KIC~10841730 in that presents only a single AB-member candidate whose $\nu_{\rm max}^{\rm spec}$, based on APOGEE data, lies between the asteroseismic $\nu_{\rm max}$ values of the two components. However, this star has a {\tt RUWE} value that indicates possible multiplicity. No further signs of activity or binarity have been found \citep[see][]{Gaulme20}. 
      
      Given that there are no additional AB-member candidates in the TPFs for these last two systems, we propose that the observed oscillations originate from spatially unresolved binaries, with the identified AB-member candidates corresponding to the primary components. Spectroscopic follow-up is needed to confirm this scenario.

      Figure~\ref{fig:binary_candidates} summarises the final sample of asteroseismic binaries identified as likely gravitationally bound systems, based on wide binary probabilities, RUWE values, and radial velocity variability indicators. Asteroseismic binaries without effective temperature estimates for lack of a \emph{Gaia} match, are assigned a $T_{\rm eff}$ around $5200-5600$~K.

\section{Summary and conclusions}\label{sec:summary}

   Asteroseismic binaries (ABs) offer a unique opportunity to detect binary systems with oscillating components and combine asteroseismology and orbital dynamics to constrain stellar properties. When gravitationally bound, these systems serve as valuable benchmarks for testing and calibrating asteroseismic scaling relations.

   In this study, we investigated the binarity of 40 seismically resolved asteroseismic binaries and used a modified version of the TACO pipeline (TACO, Hekker et al. in prep) to extract global asteroseismic parameters and derive stellar masses and radii from scaling relations calibrated with updated reference values. For each asteroseismic binary, we identified AB-member candidates from \emph{Gaia}~DR3 sources located within their \emph{Kepler} Target Pixel File (TPF). We filtered out main-sequence contaminants via CMD-based cuts using \emph{Gaia}~DR3 and cross-matched red-giants with \emph{Gaia} sources based on spectroscopic estimates of $\nu_{\rm max}^{\rm spec}$ derived from APOGEE, GSP-Spec, and GSP-Phot atmospheric parameters (the latter corrected for known systematics).

   To assess whether AB-member candidates are gravitationally bound, we applied the wide binary detection method of \citet{ElBadry21}, using \emph{Gaia} astrometry alone. For each pair, we computed projected separations and orbital velocity differences from parallaxes and proper motions, and compared them to theoretical expectations for Keplerian orbits. We then used Monte Carlo sampling to account for observational uncertainties and derived binary probabilities under different assumed total system masses. 

   The main findings of our study can be summarised as follows:
   \begin{itemize} 
      \item We identified two wide binary candidates--KIC~6501237 and KIC~10094545-- for 
      which the astrometric properties are consistent with being gravitationally bound. Although only one of the two oscillating components in each AB was confidently matched to a \emph{Gaia}~DR3 source, the combination of their relative motions and asteroseismic mass estimates supports a common origin scenario. These systems have binary probabilities of $\sim$50\% and $\sim$20\%, respectively. They represent the most relevant cases of gravitationally bound asteroseismic binaries in our sample, and they are ideal targets for future spectroscopic confirmation.

      \item {Among the 15 asteroseismic binaries where we confidently matched both oscillating components to \emph{Gaia} sources, none show astrometric evidence of gravitational binding. This indicates that most of the asteroseismic binaries in our sample are likely chance alignments rather than true binary systems. However, it's important to note that our sample is limited to seismically resolved asteroseismic binaries, and this selection bias may influence our conclusions. To reconcile our findings with the predictions from \citet{Miglio14}, future studies should systematically search for asteroseismic binaries in \emph{Kepler} long-cadence data, especially seismically unresolved systems, which \cite{Miglio14} suggest are the most common type of asteroseismic binary.}

      \item For KIC~2568888, our astrometric analysis does not allow us to confirm or rule out binarity, primarily due to an incomplete \emph{Kepler-Gaia} match. However, its only AB-member candidate has a high {\tt RUWE} value ($>1.4$), suggesting a possible unresolved companion. This system remains a promising target for future follow-up observations.

      \item We examined \emph{Gaia} multiplicity indicators (e.g. {\tt RUWE} and radial velocity variability) for additional clues on binarity. These provided limited additional information. One of our targets, KIC~7729396, shows marginal signs of RV variability, yet we were not able to draw any firm conclusions on its multiplicity. In contrast, we found that the only AB-member candidate of KIC~5443536 is most likely a spatially unresolved multiple system. Furthermore, We found five additional cases with highly significant radial velocity variability, despite presenting low {\tt RUWE} estimates. 
      
      \item Only a single AB-member candidate was found on the target pixel files of the systems KIC~6441499, KIC~8479383, and KIC~9412408. In all cases, the $\nu_{\rm max}^{\rm spec}$ values were between the two asteroseismic  $\nu_{\rm max}$ estimates. These systems resemble the case of KIC~10841730, which was confirmed as a binary in previous works. We propose that these can also be spatially unresolved binaries and encourage future spectroscopic follow-up.
   \end{itemize}

   We conclude that most asteroseismic binaries in our sample are likely chance alignments. However, two systems, KIC~6501237 and KIC~10094545, show astrometric and asteroseismic properties consistent with being wide binaries. Furthermore, we propose KIC~2568888, KIC~5443536, KIC~6441499, KIC~8479383, and KIC~10841730 as spatially unresolved binary candidates. While these represent promising candidates, further spectroscopic follow-up is required to confirm their binarity and to refine the source identification. Such observations will be essential to fully assess their potential as benchmarks for stellar and asteroseismic studies.

\begin{acknowledgements}
   We acknowledge funding from the ERC Consolidator Grant DipolarSound (grant agreement \#101000296). This paper includes data collected by the \emph{Kepler} mission and obtained from the MAST data archive at the Space Telescope Science Institute (STScI). Funding for the \emph{Kepler} mission is provided by the NASA Science Mission Directorate. STScI is operated by the Association of Universities for Research in Astronomy, Inc., under NASA contract NAS 5–26555. This work has made use of data from the European Space Agency (ESA) mission Gaia (https://www.cosmos.esa.int/gaia), processed by the Gaia Data Processing and Analysis Consortium (DPAC, https://www.cosmos.esa.int/web/ gaia/dpac/consortium). Funding for the DPAC has been provided by national institutions, in particular the institutions participating in the Gaia Multilateral Agreement.
\end{acknowledgements}

\bibliographystyle{aa}
\bibliography{bibliography}

\begin{appendix}
   \section{Peakbagging and mode identification}\label{sec:appendix_methods}
   
      In this section, we provide a brief overview of how the oscillation modes are detected, fitted, and assigned a spherical degree $\ell$ using the TACO pipeline. For a detailed description of the full methodology, we refer the reader to Hekker et al. in prep.

      \subsection{Peak detection}\label{subsec:appendix_peaks}
         To detect significant peaks around $\nu_{\text{max, A}}$ and $\nu_{\text{max, B}}$, TACO implements a Mexican hat wavelet filter following the method described by \cite{GSOM18}. The resulting peaks are then fitted using two different functions, depending on their spectral resolution $\delta\nu$. Resolved peaks are fitted using a Lorentzian profile:
            \begin{equation}
            P_{\text{peak}} (\nu) = \frac{H_{\text{peak}}}{1 + \bigg( \frac{\nu - \nu_{\text{peak}}}{\gamma_{\text{peak}}} \bigg)^{2}},
            \end{equation}
         where $H_{\text{peak}}, \nu_{\text{peak}}$, and $\gamma_{\text{peak}}$ denote the height, central frequency, and half-width at half-maximum (HWHM) of the peak, respectively. Unresolved peaks, where $\gamma_{\rm peak} \sim \delta\nu$, are modelled with a sinc-squared function:
         \begin{equation}
         \text{sinc}^{2}(\nu) = \frac{\sin{\nu}^{2}}{\nu^{2}}.
         \end{equation}
         
         Peak parameters are then optimised using a two-step fitting process. First, TACO performs a global maximum likelihood fit using the parameters derived from the detection process as initial guess. Subsequently, each peak is fitted individually while keeping the rest of the model fixed, and the uncertainties are estimated from the diagonal elements of the inverse of the Hessian matrix.
            
      \subsection{Mode identification}\label{subsec:appendix_modeID}
         TACO identifies the central radial mode closest to $\nu_{\rm max}$ by cross-correlating the observed PDS with the universal pattern of $\ell=0$ solar-like oscillations \citep{Mosser11}
         \begin{equation}
         \nu_{n,\ell=0} = \bigg( n + \epsilon_p + \frac{\alpha}{2} (n-n_{\rm max})^{2} \bigg)\Delta\nu,
         \end{equation}
         where $\epsilon_{\rm p}$ is the phase offset, $\alpha$ the curvature term, and $n_{\rm max} = \nu_{\rm max}/\Delta\nu$ the radial order nearest to $\nu_{\rm max}$. We used $\nu_{\rm max}$ from the background fit and $\Delta\nu$ from the scaling relation of \cite{Hekker11} as an initial estimate. Subsequently, TACO identifies radial modes at higher and lower radial orders based on this relation.
            
         Using the frequencies of all the significant $\ell=0$ modes, TACO performs a linear regression to derive global estimates of the seismic parameters $\Delta\nu$, $\epsilon_{\rm p}$, and $\alpha$, as well as central values, $\Delta\nu_{c}$, $\epsilon_{\rm p,c}$, and $\alpha_{c}$ from the three central radial modes.

         Finally, TACO searches for quadrupole ($\ell=2$) and octupole ($\ell=3$) modes. These are identified as broad peaks near the expected frequency region, based on estimates of $\delta\nu_{02}$ and $\delta\nu_{03}$ from \cite{Huber10} and \cite{Mosser11UP}, respectively.
         
         This version of TACO does not identify dipole modes ($\ell=1$), which often exhibit complex mixed-mode patterns in red giants. These modes result from coupling between pressure and gravity waves, leading to deviations from the regular frequency spacing found in pure pressure modes, i.e. radial modes. Their identification requires more sophisticated modelling. Here, all significant peaks not assigned to $\ell=0,2,$ and $3$ are flagged as dipole-mode candidates.

   \section{Evolutionary stage determination}\label{sec:appendix_ES}
\begin{figure}
         \center
         \includegraphics[width=0.45\textwidth]{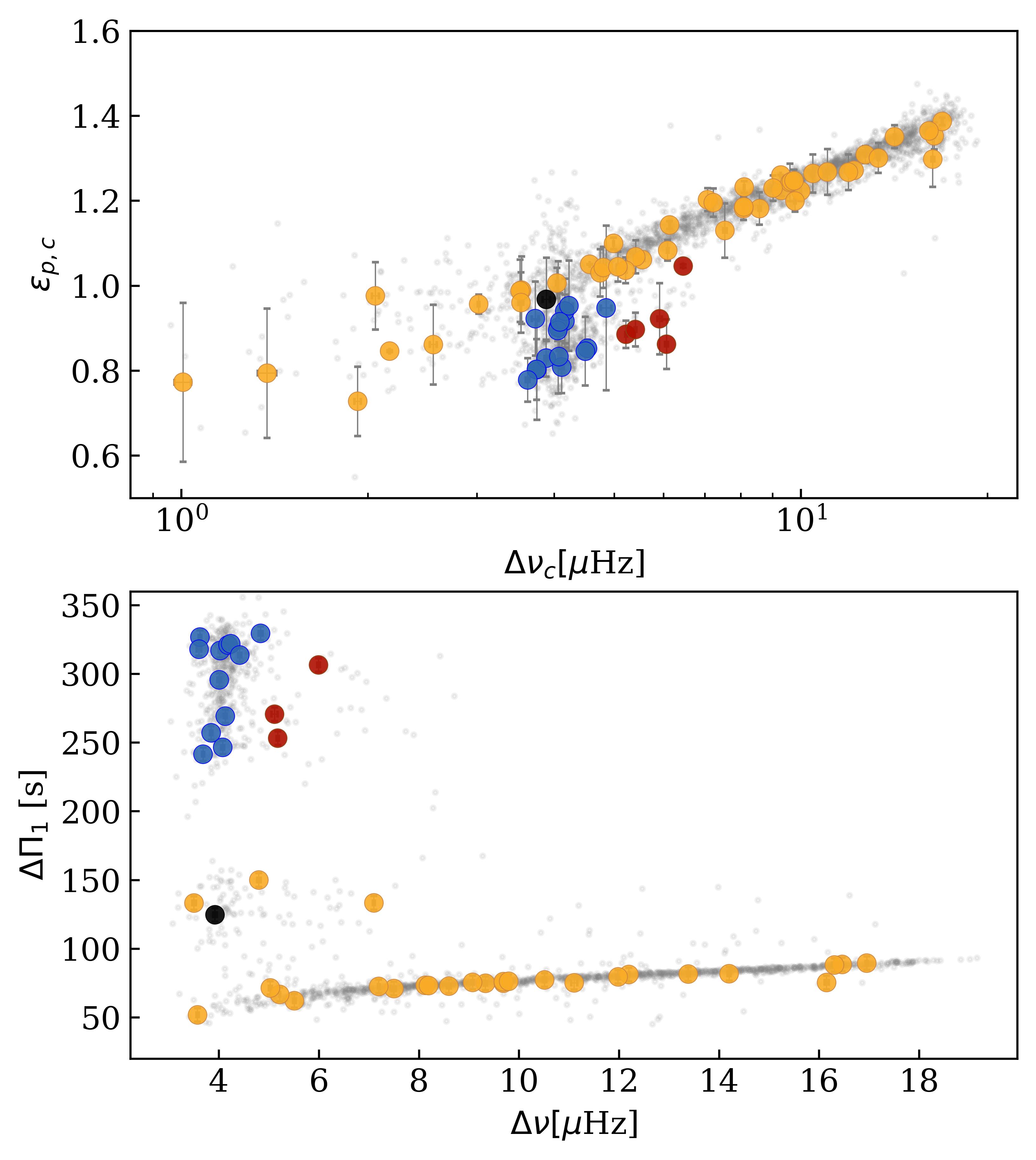} 
         \caption{Evolutionary stages of our sample. Top: Phase shift of the three central radial modes, $\epsilon_{c}$ as a function of the central large frequency separation $\Delta\nu_{c}$. Bottom: period spacing $\Delta\Pi_{1}$ as a function of $\Delta\nu$. The stars are colour-coded by their evolutionary stage. Orange, blue and red circles are red-giant branch (RGB), red clump (RC), and secondary clump (2CL), respectively. A star with ambiguous evolutionary stage is shown as a black dot. The grey-background circles correspond to the "golden sample" from CAPASS (Espinoza-Rojas et al. in prep).} \label{fig:evol_stages}
      \end{figure}

      The evolutionary stage of red-giant stars can be determined from asteroseismic parameters. As shown by \cite{Kallinger12}, the structural differences between red-giant branch (RGB) and red clump (RC) stars affect parameters derived from pure pressure modes, i.e. $\ell=0$. In particular, the central estimates of the large frequency separation $\Delta\nu_{c}$ and phase offset $\epsilon_{\rm p,c}$ are sensitive to these differences: core-helium burning stars typically exhibit lower values of $\epsilon_{p,c}$ than hydrogen-shell burning stars at the same $\Delta\nu_{c}$, as illustrated in the top panel of Fig.~\ref{fig:evol_stages}. As discussed in \cite{Kallinger12}, red clump stars have a narrow range in $\Delta\nu_c$ due to their common core mass at the helium flash, while secondary clump stars, which are more massive, ignite helium non-degeneratively, allowing for a range of core masses and $\Delta\nu_c$ values. However, there is a region in the $\Delta\nu_{c}-\epsilon_{\rm p,c}$ space where RGB and RC stars overlap, complicating their classification using this method alone.

      In such cases, the presence of dipole mixed-modes is valuable, since they provide a more direct probe of the stellar core and hence the evolutionary stage. These modes exhibit a mixed character in red-giant stars, behaving as pressure-modes ($p$-modes) in the stellar envelope and as gravity-modes ($g$-modes) in the core due to coupling between the respective mode cavities \citep{Bedding14}. Mixed-modes follow a distinct pattern characterised by a near constant separation in period, also known as the period spacing $\Delta\Pi_{l}$ \citep{Tassoul1980}. This parameter carries indirect information about the core density and in dipole modes reaches values above $200$~s for core-helium burning stars, while RGB stars show values between 50 and 100~s \citep[e.g.][]{Mosser15,Vrard16}. This separation enables clear classification in the $\Delta\nu-\Delta\Pi_{1}$ diagram \citep[e.g.][]{Bedding11,Mosser11, Mosser12, Mosser14} (Fig.~\ref{fig:evol_stages}, bottom panel),

      We estimated $\Delta\Pi_{1}$ by computing the Lomb-Scargle periodogram of the PDS without the $\ell=0,2$ and $3$ modes. However, our sample of stars are affected by photometric dilution, where flux blending from nearby sources in the same aperture reduces the observed oscillation amplitudes \citep[e.g.][]{Themessl18a}, making the mode identification challenging and reducing the signal of modes with long lifetimes, i.e. narrow peaks. In such cases, the reliable estimation of $\Delta\Pi_{1}$ is hampered. To mitigate this, we visually inspected the PDS to confirm whether the assigned evolutionary stage was consistent with the observed mode density, particularly the complex mode patterns expected in RC stars. The final evolutionary stages are shown in the lower panel of Fig.~\ref{fig:evol_stages}.

      As illustrated in Fig.~\ref{fig:evol_stages}, one target remains ambiguous. Its position in both the $\Delta\nu_{c}-\epsilon_{\rm p,c}$ and $\Delta\nu-\Delta\Pi_{1}$ diagrams lies between the RGB and RC regions. Similar cases have not been reported yet, but our analysis of single stars in CAPASS (Espinoza-Rojas et al. in prep) combined with evolutionary stages from \cite{Elsworth19} suggest that both RGB and RC stars may appear in this region, and we cannot discard this arising from biases in the determination of $\Delta\Pi_{1}$.

   \section{Limitations of wide binary detection for $\varpi < 1$~mas}\label{sec:appendix_wb}
\begin{figure}
         \center
         \includegraphics[width=0.45\textwidth]{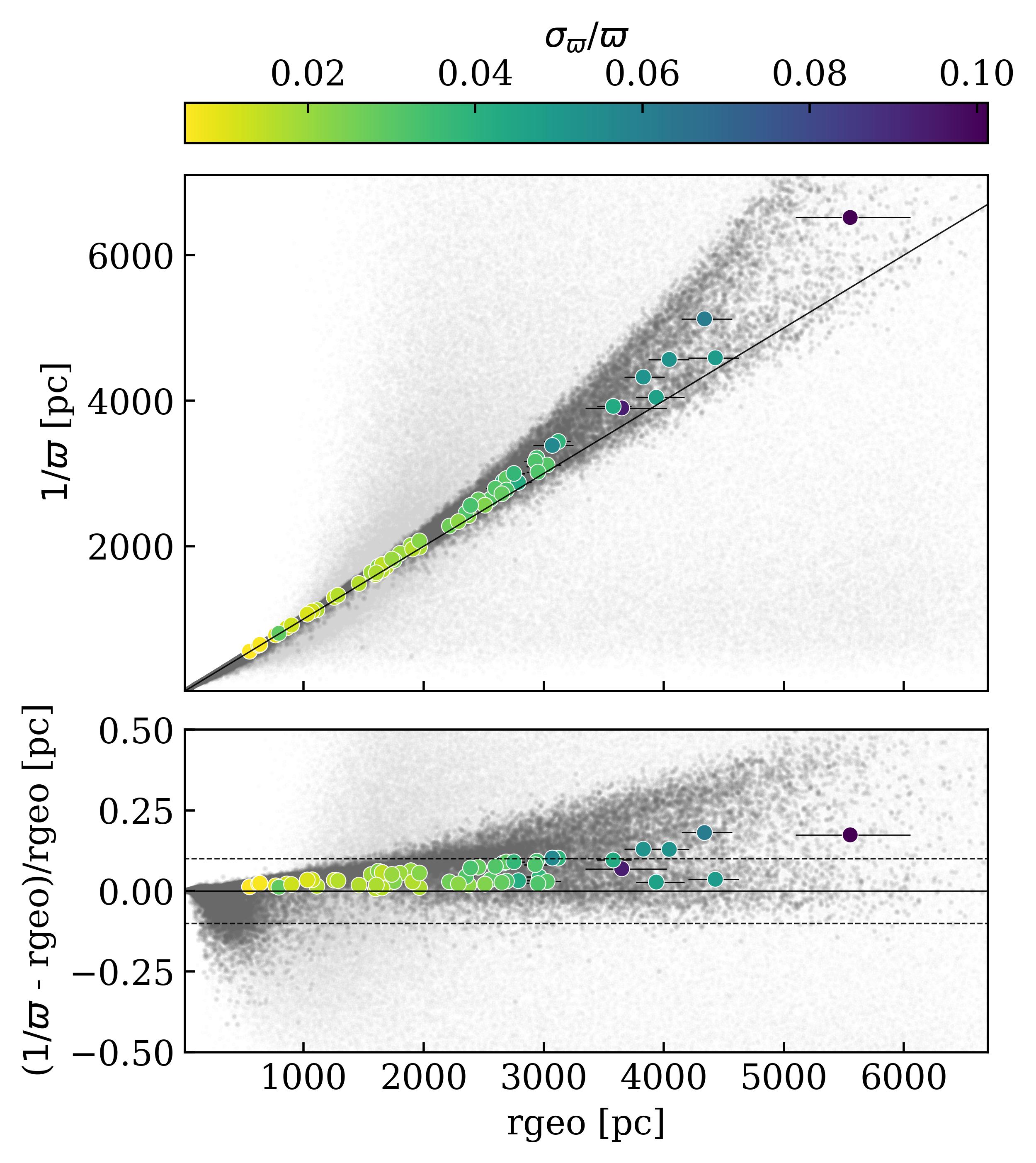} 
         \caption{Comparison of \emph{Gaia}~DR3 inverse parallaxes as a function of geometric distances from \cite{BJ21}. Dark grey background points represent a sample of 40000 randomly selected \emph{Gaia}~DR3 stars with fractional parallax error $\sigma_{\varpi}/\varpi<0.2$. Light grey points show stars with $\sigma_{\varpi}/\varpi>0.2$ from \cite{BJ21}. Our sample of AB-member candidates are shown as circles, colour coded by their fractional parallax error. The bottom panel shows the respective residuals. Black dashed lines depict a difference of $\pm10\%$.} \label{fig:plx_distance}
      \end{figure}

      As discussed in \cite{BJ15}, the use of inverse parallax as distance proxy ($d\sim1/\varpi$) is only reliable in the absence of noise. When the fractional parallax uncertainty exceeds $20\%$ ($\sigma_{\varpi}/\varpi>0.2$), a probabilistic inference approach is required to obtain an unbiased estimate of distance. In this study, we applied quality cuts to exclude stars with large parallax uncertainties, thereby restricting our analysis to sources where the inverse-parallax remains valid as a proxy (dark grey points in Fig.~\ref{fig:plx_distance}). To evaluate the impact of this assumption, we compared the inverse-parallax distances with those inferred by \cite{BJ21} from a probabilistic model. 
      
      As illustrated in Fig.~\ref{fig:plx_distance}, the use of inverse parallax as a distance proxy is generally reliable for stars with good parallax quality, i.e. $\sigma_{\varpi}/\varpi < 0.2$, up to distances of approximately 2~kpc. Beyond this range, the proxy tends to overestimate distances by $\geq$10\%. The light grey background in Fig.~\ref{fig:plx_distance} represent stars with large fractional parallax uncertainties, illustrating that, without accounting for parallax quality, significant distance overestimations can arise already at $\sim$1~kpc. Consequently, the use of inverse parallax in the context of wide binary detection can introduce systematic errors in the derived projected separations and tangential velocity differences, potentially affecting the estimated binarity likelihood.

      This limitation is particularly relevant for red-giant stars, which are intrinsically luminous and thus frequently observed at distances greater than 1~kpc. Careful consideration must therefore be given when applying the wide-binary detection method based on total orbital velocity differences to these stars, as the assumptions underlying the method can lead to ambiguous conclusions.

      To quantify the impact of the inverse parallax distance prior on our binarity likelihood, we recomputed projected separations, tangential velocities and total orbital velocity differences for our whole set of AB-member candidates using the distances from \cite{BJ21}. The equations presented in Sect.~\ref{sec:grav_bound} were modified to use the distances instead of $1/\varpi$. Hence, the projected separation becomes:
      \begin{equation}
         s = \bigg(\frac{\theta}{\text{arcsec}} \bigg) \bigg(\frac{d}{\text{pc}} \bigg) {\rm AU},
      \end{equation}
      whereas Eq.~\ref{eq:max_s} can be expressed as:
      \begin{equation}\label{eq:max_s_2}
         \frac{\theta}{\text{arcsec}} \leq 206265 \times \bigg( \frac{d}{\rm pc} \bigg)^{-1}.
      \end{equation}
      And finally, the tangential velocities can be rewritten as
      \begin{equation}
         \Delta V_{\perp} = 4.74 \times \bigg( \frac{\Delta\mu}{\text{arcsec yr}} \bigg)\times \bigg( \frac{d}{\rm pc}\bigg) \text{ km s}^{-1}
      \end{equation}
      
      Note that the parallax consistency condition (Eq.~\ref{eq:parallax_diff}) cannot be straightforwardly reformulated in terms of distance, as the interpretation of the uncertainty in parallax does not translate directly to distance space. The purpose of this exercise is merely to assess how the orbital velocities change when geometric distance estimates are used, rather than to redefine the consistency criterion. A full derivation of a revised distance-based condition is beyond the scope of this work. Furthermore, replacing inverse parallaxes with distance estimates from \citet{BJ21} leads to an increase in the uncertainties of the orbital velocity differences. For this reason, we do not report these revised values as final results.

\begin{figure}
         \sidecaption
         \centering
         \includegraphics[width=0.65\textwidth]{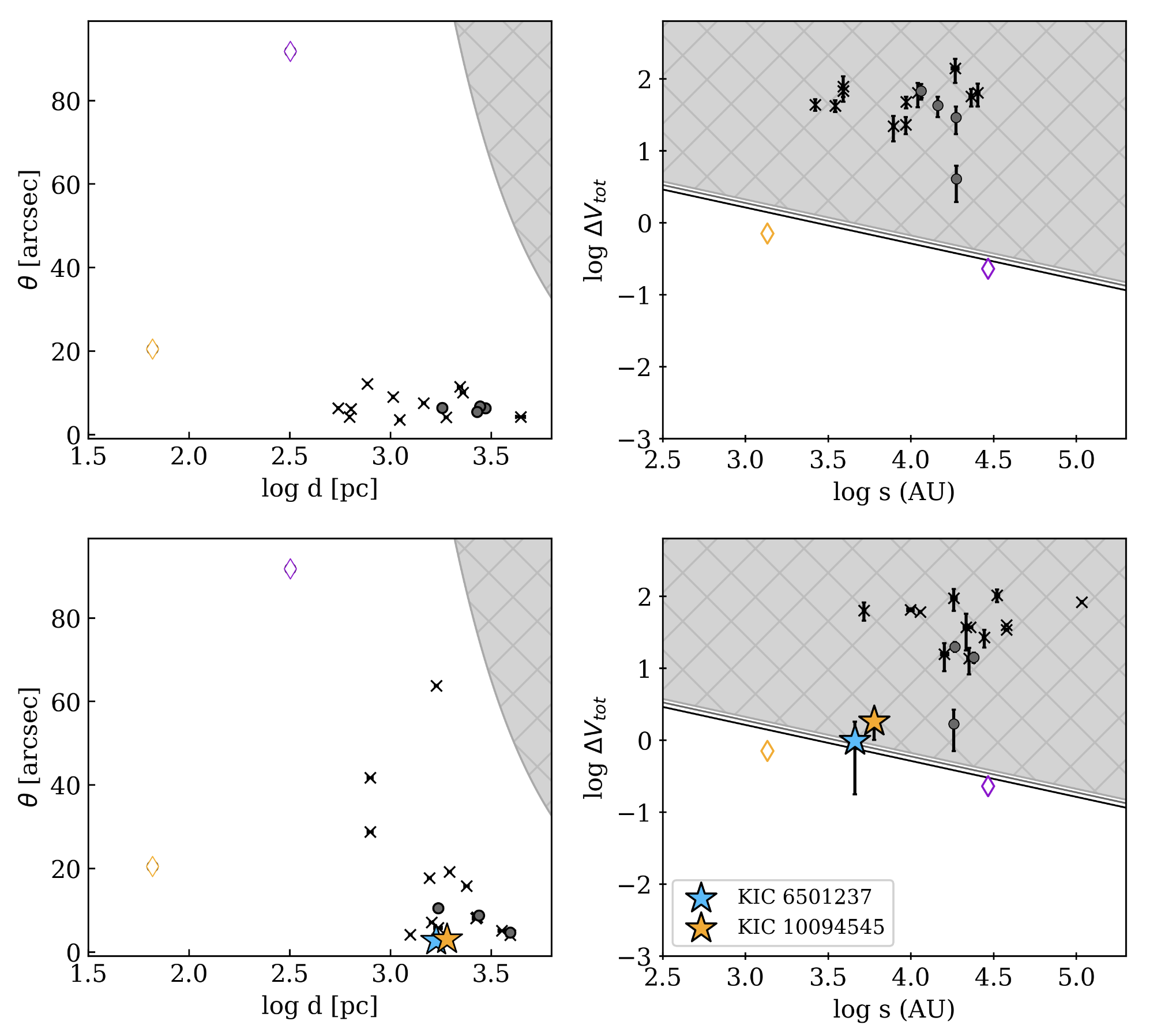}
         \caption{Same as left and right columns of Fig.~\ref{fig:grav_bound}, now using geometric distances from \cite{BJ21} instead of inverse parallaxes for the computation of projected separations and orbital velocities.}\label{fig:grav_bound_dist} 
      \end{figure}

      As shown in Fig.~\ref{fig:grav_bound_dist}, the resulting changes in binarity likelihood are negligible and do not affect the conclusions presented in Fig.~\ref{fig:grav_bound} or the main finding of this study. 

      \onecolumn

      \section{\emph{Kepler-Gaia} cross-match results}\label{sec:appendix_tables}
      
\begin{figure}[ht]
            \centering
            \begin{minipage}{0.88\textwidth}
               \centering
               \includegraphics[width=0.88\textwidth]{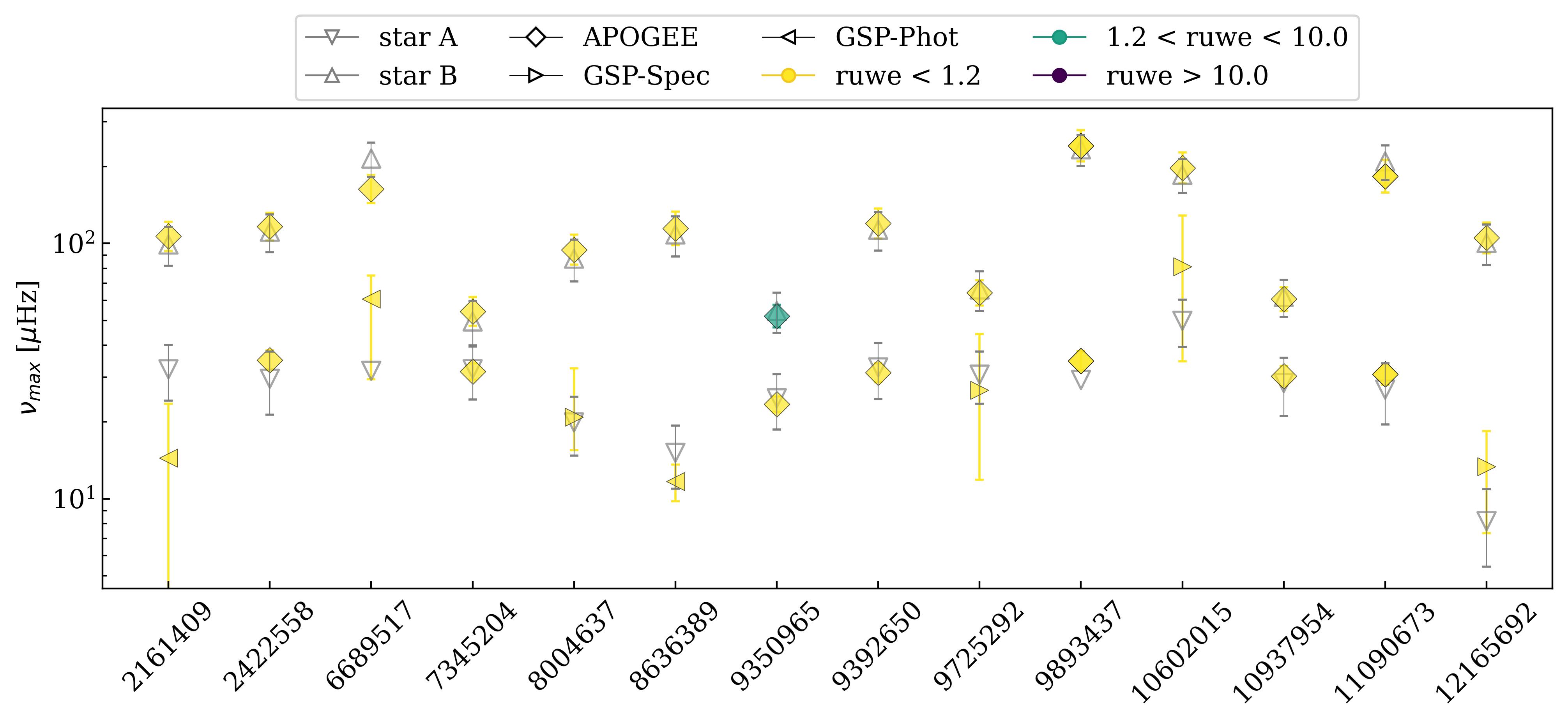}  
            \end{minipage}
            \vspace{0.01cm} 
            \begin{minipage}{0.88\textwidth}
               \centering
               \includegraphics[width=0.88\textwidth]{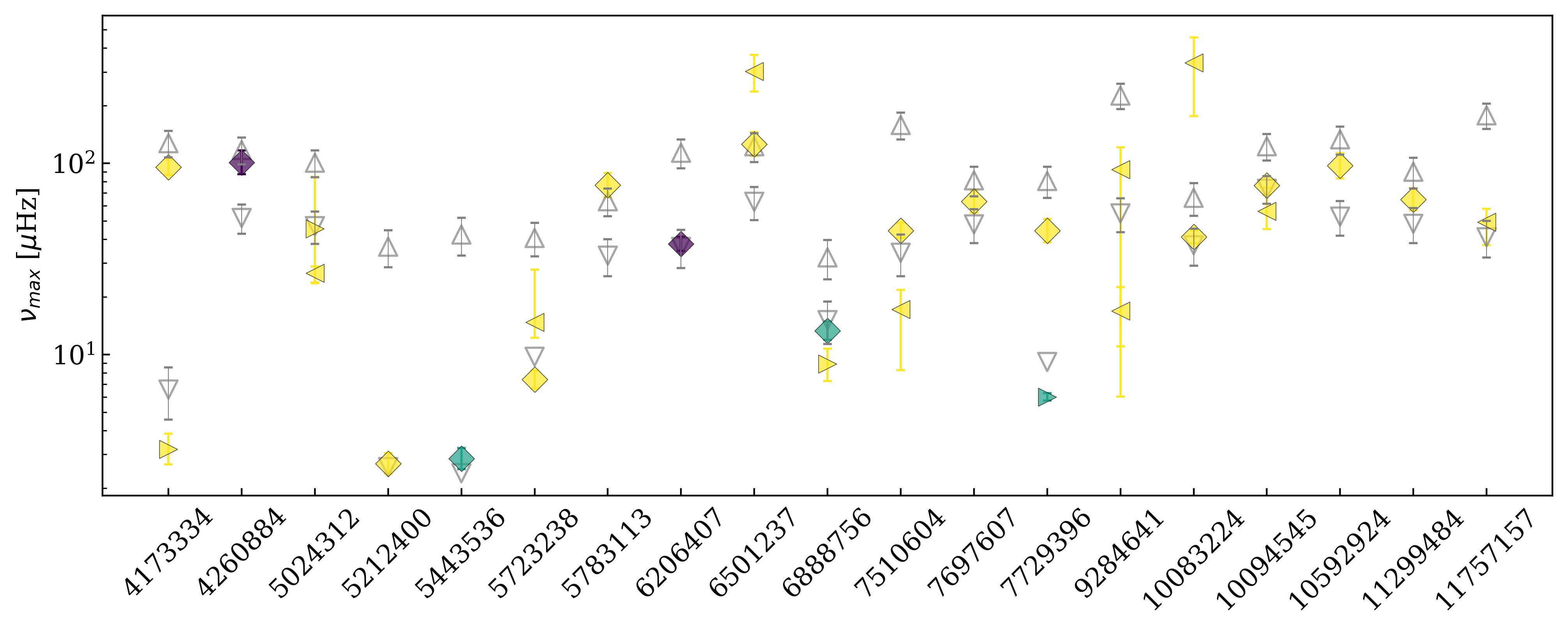}
            \end{minipage}   
            \vspace{0.01cm} 
            \begin{minipage}{0.88\textwidth}
               \centering
               \includegraphics[width=0.88\textwidth]{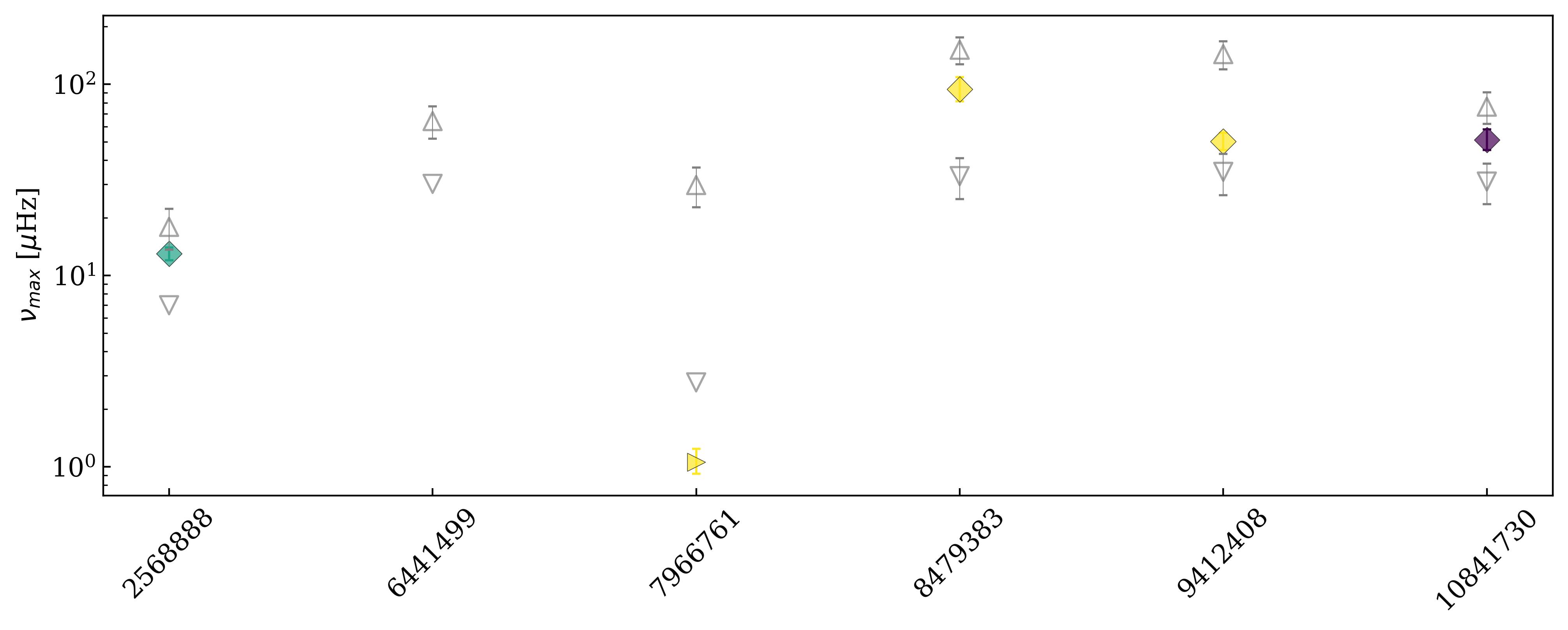} 
               \caption{Results of matching \emph{Kepler} asteroseismic binaries with \emph{Gaia}~DR3 sources. Top, centre and bottom panels show the asteroseismic binaries were both, only one and none of oscillators were matched \emph{Gaia} source(s), respectively (see Tables~\ref{table:Full_identification_summary}, ~\ref{table:single_identification_summary}, and ~\ref{table:null_identification_summary}). Grey up and down triangles indicate the asteroseismic $\nu_{\rm max}$ estimates. Diamonds, and right and left triangles represent spectroscopic $\nu_{\rm max}$ estimates derived from APOGEE, GSP-Spec and GSP-Phot stellar parameters, respectively. Stars with no indication of being unresolved multiple systems ({\tt RUWE $<1.2$}) are shown yellow, while stars with larger {\tt RUWE} values are shown in green and purple. The only AB-member candidate of KIC~6441499 was excluded from the selection of red-giant stars due to the lack of \emph{Gaia} photometry; therefore it is not shown in this plot (see Sect.~\ref{subsec:single_non_match}).}\label{fig:matching_results}  
            \end{minipage}
         
         \end{figure}

         In this section, we summarise the cross-match between \emph{Kepler} asteroseismic binaries and \emph{Gaia}~DR3 sources described in Sect.~\ref{sec:surveys_xmatch}. The tables list seismic parameters, evolutionary stages, and \emph{Gaia} multiplicity indicators (NSS solutions and {\tt RUWE}) for three cases: both stars matched, one star matched, or no match. Figure~\ref{fig:matching_results} illustrates these cases and compares seismic and spectroscopic $\nu_{\rm max}$ estimates.

      \subsection{Tables}
         \begin{table}[ht]
            \caption{List of asteroseismic parameters for systems where both solar-like oscillators were identified and matched to a \emph{Gaia} DR3 source. } 
            \label{table:Full_identification_summary}   
            \centering                     
         \begin{tabular}{ccccccccc}
            \hline\hline  
            
            KIC & $\nu_{\text{max, A}}$ &  $\nu_{\text{max, B}}$ &  M$_{\text{A}}$ &  M$_{\text{B}}$ &  ES$_{\text{A}}$ &  ES$_{\text{B}}$ & NSS$_{\text{A}}$ & NSS$_{\text{B}}$\\
               & [$\mu$Hz]             &   [$\mu$Hz]            &  [M$_{\odot}$]    &  [M$_{\odot}$]    &                  &                  \\
            \hline
            2161409 &  32.1$\pm$0.42 &   99.0$\pm$0.37 & 1.11$\pm$0.05 & 1.52$\pm$0.03 &     RGB* &     RGB &   0 &   0 \\
            2422558 & 29.49$\pm$0.32 & 111.11$\pm$0.93 & 0.78$\pm$0.05 & 1.50$\pm$0.03 &      RC  &     RGB &   0 &   0 \\
            4663623 &  34.1$\pm$1.02 &  58.66$\pm$0.91 & 1.22$\pm$0.07 & 2.26$\pm$0.04 &      RC  &     RGB &   0 &   2 \\
            6689517 & 31.67$\pm$2.56 & 214.72$\pm$0.82 &    -          & 1.13$\pm$0.02 & Unknown  &     RGB &   0 &   0 \\
            7345204 & 32.13$\pm$0.49 &  49.44$\pm$0.55 & 1.23$\pm$0.04 & 1.54$\pm$0.03 &      RC  &     RGB &   0 &   0 \\
            8004637 & 19.89$\pm$0.73 &  87.13$\pm$0.38 &   -           & 1.26$\pm$0.03 &     RGB  &     RGB &   0 &   0 \\
            8636389 & 15.13$\pm$0.39 & 108.36$\pm$0.65 & 1.31$\pm$0.06 & 1.29$\pm$0.03 &     RGB  &     RGB &   0 &   0 \\
            9350965 & 24.73$\pm$0.28 &  54.35$\pm$0.77 & 2.57$\pm$0.05 & 1.38$\pm$0.04 &     RGB  &      RC &   0 &   0 \\
            9392650 & 32.66$\pm$0.58 & 113.17$\pm$4.64 & 1.08$\pm$0.05 & 1.36$\pm$0.10 &      RC  &     RGB &   0 &   0 \\
            9725292 & 30.60$\pm$0.46 &  65.98$\pm$1.91 & 1.52$\pm$0.04 & 2.19$\pm$0.08 &     RGB  &     2CL &   0 &   0 \\
            9893437 & 33.26$\pm$0.06 & 232.91$\pm$0.88 &   -            & 1.65$\pm$0.03 & Unknown  &     RGB &   0 &   0 \\
            10602015 & 49.75$\pm$1.86 & 185.69$\pm$0.74 &  -            & 1.39$\pm$0.03 &    RGB  &     RGB &   0 &   0 \\
            10937954 & 28.37$\pm$0.81 &  61.68$\pm$3.61 & 1.12$\pm$0.07 & 3.04$\pm$0.15 &     RC  &     2CL &   0 &   0 \\
            11090673 & 26.74$\pm$1.15 & 209.35$\pm$0.56 & 0.97$\pm$0.11 & 1.13$\pm$0.03 &     RC  &     RGB &   0 &   0 \\
            12165692 &  8.16$\pm$0.20 & 100.44$\pm$0.28 & 1.13$\pm$0.07 & 1.28$\pm$0.03 &    AGB  &     RGB &   0 &   0 \\
         
            \hline
         \end{tabular}
         \end{table}
         
         \begin{table}[h]
            \caption{Asteroseismic parameters for systems which only one of the two solar-like oscillators was identified and matched to a \emph{ Gaia} DR3 source.} \label{table:single_identification_summary}   
            \centering                     
         \begin{tabular}{ccccc}
            \hline\hline  
            KIC &        $\nu_{\rm max}$ &          $M$  &      ES  &  {\tt RUWE} \\
               &          [$\mu$Hz]        & [M$_{\odot}$] &          &             \\ 
            \hline \ 
            4173334A & 6.56$\pm$0.19   &               &     RGB &          \\
            4173334B & 127.18$\pm$8.25 & 1.48$\pm$0.16 &     RGB &     1.00 \\
            \hline
            4260884A &  51.68$\pm$0.78 &               &      RC &          \\
            4260884B & 117.17$\pm$1.85 & 1.53$\pm$0.06 &     RGB &    13.04 \\
            \hline
            5024312A &  46.85$\pm$1.23 & 2.00$\pm$0.07 &     RGB &     1.01 \\
            5024312B & 100.54$\pm$1.23 &               &     RGB &          \\
            \hline
            5212400A &   2.58$\pm$0.01 & 1.14$\pm$0.24 &     RGB &     1.04 \\
            5212400B &  36.59$\pm$0.91 &               &      RC &          \\
            \hline
            5443536A &   2.39$\pm$0.08 & 1.14$\pm$0.25 &     RGB &     1.36 \\
            5443536B &  42.34$\pm$2.90 &               &     RGB &          \\
            \hline
            5723238A &   9.72$\pm$0.63 & 2.05$\pm$0.12 &     RGB &     1.04 \\
            5723238B &  40.63$\pm$2.24 &               &     RGB &          \\
            \hline 
            5783113A &  32.84$\pm$2.16 &               &     RGB &          \\
            5783113B &  63.21$\pm$4.37 & 2.99$\pm$0.16 &     2CL &     1.18 \\
            \hline 
            6206407A &  36.50$\pm$0.39 & 1.41$\pm$0.04 &      RC &    34.04 \\
            6206407B & 113.37$\pm$4.70 &               &     RGB &          \\
            \hline 
            6501237A &  62.78$\pm$0.29 &               &     RGB &          \\
            6501237B & 122.54$\pm$1.09 & 1.27$\pm$0.03 &     RGB &     1.03 \\
            \hline 
            6888756A &  15.11$\pm$0.25 & 1.87$\pm$0.04 &     RGB &     1.81 \\
            6888756B &  32.16$\pm$0.25 &               &      RC &          \\
            \hline 
            7510604A &  34.00$\pm$0.67 & 1.03$\pm$0.05 &      RC &     1.10 \\
            7510604B & 158.32$\pm$3.40 &               &     RGB &          \\
            \hline 
            7697607A &  47.84$\pm$0.60 &               &     RGB &          \\
            7697607B &  81.34$\pm$4.28 & 1.72$\pm$0.04 &     RGB &     0.95 \\
            \hline 
            7729396A &   9.13$\pm$0.52 & 1.36$\pm$0.10 &     RGB &     3.72 \\
            7729396B &  80.69$\pm$3.77 &               &     RGB &          \\
            \hline
            9284641A &   54.61$\pm$0.25 & 1.16$\pm$0.03 &    RGB &     0.98 \\
            9284641B &  225.53$\pm$0.80 &               &    RGB &          \\
            \hline 
            10083224A &  37.19$\pm$0.31 & 1.53$\pm$0.04 &     RC &     0.98 \\
            10083224B &  65.88$\pm$7.17 &               &     RC &          \\
            \hline 
            10094545A &  73.41$\pm$0.56 & 2.06$\pm$0.12 &    2CL &     1.12 \\
            10094545B & 122.55$\pm$4.27 &               &    RGB &          \\
            \hline 
            10592924A &  52.54$\pm$0.82 &               &    RGB &          \\
            10592924B & 133.21$\pm$2.11 & 1.35$\pm$0.05 &    RGB &     1.11 \\
            \hline 
            11299484A &  48.18$\pm$0.29 &               &    RGB &          \\
            11299484B &  90.19$\pm$0.48 & 1.42$\pm$0.03 &    RGB &     1.09 \\
            \hline 
            11757157A &  40.97$\pm$0.34 & 1.37$\pm$0.09 &     RC &     0.93 \\
            11757157B & 177.86$\pm$1.05 &               &    RGB &          \\
            \hline
         \end{tabular}
         \end{table}

         \begin{table}[h]
            \caption{Asteroseismic parameters for systems in which neither of the two solar-like oscillators was matched to a \emph{Gaia} DR3 source.} \label{table:null_identification_summary}   
            \centering                     
            \begin{tabular}{ccccc}
               \hline\hline  
                  KIC & $\nu_{\text{max, A}}$ & $\nu_{\text{max, B}}$ & ES$_{\text{A}}$ & ES$_{\text{B}}$ \\
                     &         [$\mu$Hz]     &     [$\mu$Hz]         &                 &                  \\ 
               \hline
                  2568888  &   6.98$\pm$0.34 &   17.94$\pm$0.44 &     RGB &  RGB \\
                  6441499  &  30.08$\pm$0.42 &   64.12$\pm$1.56 & Unknown &  RGB \\
                  7966761  &   2.76$\pm$0.04 &   29.69$\pm$0.27 &     RGB &  RGB \\
                  8479383  &  33.03$\pm$0.43 &  151.39$\pm$2.72 &      RC &  RGB \\
                  9412408  &  34.77$\pm$0.52 &  143.60$\pm$3.08 &      RC &  RGB \\
                  10841730 &  30.94$\pm$0.36 &   76.27$\pm$3.04 &      RC &  RGB \\
               
               \hline
            \end{tabular}
         \end{table}
      
      \clearpage      
      \subsection{Figures}

   \end{appendix}

\end{document}